\documentclass{aa}

\usepackage{graphicx}
\usepackage{subcaption}
\usepackage{color}

\usepackage{txfonts}
\usepackage[colorlinks=true,linkcolor=blue,filecolor=blue,citecolor=blue,urlcolor=blue]{hyperref}
\begin{document}

   \title{Timing irregularities and glitches from the pulsar monitoring campaign at IAR}


   \author{E. Zubieta
          \inst{1,2}\fnmsep\thanks{E-mail: ezubieta@iar.unlp.edu.ar}
          \and
          F. Garc\'{i}a\inst{1,2}
          \and
          S. del Palacio\inst{1,3}
          \and
          S. B. Araujo Furlan\inst{4,5}
          \and
          G. Gancio\inst{1}
          \and
          C. O. Lousto\inst{6,7}
          \and
          J. A. Combi\inst{1,2}
          \and
          C. M. Espinoza\inst{8,9}
          }

   \institute{Instituto Argentino de Radioastronom\'ia (CCT La Plata, CONICET; CICPBA; UNLP),
              C.C.5, (1894) Villa Elisa, Buenos Aires, Argentina
         \and
             Facultad de Ciencias Astron\'omicas y Geof\'{\i}sicas, Universidad Nacional de La Plata, Paseo del Bosque, B1900FWA La Plata, Argentina
         \and
             Department of Space, Earth and Environment, Chalmers University of Technology, SE-412 96 Gothenburg, Sweden
         \and
             Instituto de Astronom\'\i{}a Teórica y Experimental, CONICET-UNC, Laprida 854, X5000BGR – Córdoba, Argentina
         \and
             Facultad de Matemática, Astronomía, Física y Computación, UNC. Av. Medina Allende s/n , Ciudad Universitaria, CP:X5000HUA - Córdoba, Argentina.
         \and
             School of Mathematical Sciences, Sciences Rochester Institute of Technology Rochester, NY 14623, USA
         \and
             Center for Computational Relativity and Gravitation, Rochester Institute of Technology, 85 Lomb Memorial Drive, Rochester, New York 14623, USA
        \and Departamento de F\'isica, Universidad de Santiago de Chile (USACH),  Av. V\'ictor Jara 3493, Estaci\'on Central, Chile.
        \and Center for Interdisciplinary Research in Astrophysics and Space Sciences (CIRAS), Universidad de Santiago de Chile.
             }

   \date{Received ...; accepted ...}

 
  \abstract
   {Pulsars have an overall very stable rotation. However, sudden increases in their rotation frequency, known as glitches, perturb their evolution. While many observatories commonly detect large glitches, small glitches are harder to detect because of the lack of daily-cadence observations over long periods of time (years).}
   {We aim to explore and characterise the timing behaviour of young pulsars at daily timescales, looking for small glitches and other irregularities. The results will further our comprehension of the real distribution of glitch sizes, which has also consequences for the theoretical modeling of the glitch mechanism.}
   {We observed six pulsars with up to daily cadence between December 2019 and January 2024 with the two antennas of the Argentine Institute of Radio Astronomy (IAR). We used standard pulsar timing tools to obtain the times of arrival of the pulses and to characterise the pulsar's rotation. We developed an algorithm to look for small timing events in the data and calculate the changes in the frequency ($\nu$) and its derivative ($\dot\nu$) at those epochs.} 
   {We found that the rotation of all pulsars in this dataset is affected by small step changes in $\nu$ and $\dot \nu$. Among them, we found three new glitches that had not been reported before: two glitches in PSR~J1048$-$5832 with relative sizes $\Delta\nu / \nu=9.1(4)\times 10^{-10}$ and $\Delta\nu / \nu=4.5(1)\times 10^{-9}$, and one glitch in the Vela pulsar with a size $\Delta\nu / \nu=2.0(2)\times 10^{-10}$. We also report new decay terms on the 2021 Vela giant glitch, and on the 2022 giant glitches in PSR J0742$-$2822 and PSR J1740$-$3015 respectively. In addition, we found that the red noise contribution significantly diminished in PSR J0742$-$2822 after its giant glitch in 2022.}
   {Our results highlight the importance of high-cadence monitoring with an exhaustive analysis of the residuals to better characterize the distribution of glitch sizes and to deepen our understanding of the mechanisms behind glitches, red noise and timing irregularities.}

   \keywords{(Stars:) pulsars: general -- 
                 methods: observational  --
                radio continuum: general
               }

   \maketitle
%

\section{Introduction}

   Pulsars are neutron stars that exhibit pulsed emission, primarily in radio frequencies. Their extremely high moment of inertia provides them with exceptionally stable rotation, which positions some pulsars among the most stable clocks in the universe \citep{2012MNRAS.427.2780H}. If a pulsar is isolated, the evolution of its rotation is rather smooth, decelerating steadily as a result of energy loss via dipole emission. 
   However, dynamic effects may perturb this evolution, primarily in young pulsars. These effects are known as glitches and timing noise \citep{2004hpa..book.....L}. 
   
   Glitches are discrete events consisting of a sudden change in rotation frequency, most of the times accompanied by sudden changes in the frequency derivative, which typically returns to its pre-glitch value exponentially \citep{Zhou2022}. At present, more than 200 pulsars have presented glitches\footnote{http://www.jb.man.ac.uk/pulsar/glitches.html} \citep{2011MNRAS.414.1679E, 2017A&A...608A.131F,2018IAUS..337..197M,2022MNRAS.510.4049B}. These are believed to be caused by the decoupling between the superfluid interior of the star and the solid crust \citep{1969Natur.224..673B, 2015IJMPD..2430008H, 2022MNRAS.511..425G}. The magnitude of the glitches provides insight into the reservoir of angular momentum accessible in the superfluid interior \citep{2022Univ....8..641Z}. 
   Glitch sizes, characterised by the relative jump in their spin frequency ($\Delta \nu / \nu$), exhibit a bimodal distribution \citep{2011MNRAS.414.1679E, 2013MNRAS.429..688Y}, which separates glitches \citep{2017A&A...608A.131F} into giant glitches ($\Delta \nu / \nu \sim 10^{-6}$) and small glitches ($\Delta \nu / \nu \sim 10^{-9}$).
   
   On the other hand, timing noise manifests itself as a smooth and wandering behaviour around a simple rotational evolution, and its explanation remains a puzzle \citep{2010MNRAS.402.1027H, 2019MNRAS.489.3810P}. It is unknown whether timing noise may have similar causes, such as superfluid turbulence \citep{2009ApJ...700.1524M} or unresolved microglitches \citep{2006A&A...457..611J}, stem from entirely different mechanisms, such as fluctuations of the magnetosphere \citep{2010Sci...329..408L}, or arise from the combination of both the internal superfluid torque and external magnetospheric torque \citep{2023MNRAS.520.2813A}.
   
   Although the distinction between timing noise and glitches is evident in terms of the discreteness of the event \citep{2016MNRAS.459.3104S}, the smallest glitches can often be mistaken for timing noise and vice versa \citep{2024arXiv240514351G}. In particular, even though low sensitivity can also obscure detections, the main cause of the confusion between a glitch-like event and timing noise is that the monitoring cadence is often too low to evaluate the discreteness of the event. Therefore, by exploring the distribution of small irregularities in pulsar timing with high-cadence observations, more information can be obtained about the connection or difference between glitches and timing noise.
   
    Our Pulsar Monitoring in Argentina\footnote{\url{https://puma.iar.unlp.edu.ar}} (PuMA) collaboration has been observing a set of bright pulsars in the southern hemisphere since 2019 \citep{2020A&A...633A..84G,2024RMxAC..56..134L} using the 30-m antennas of the the Argentine Institute of Radioastronomy (IAR). In particular, we have been closely monitoring the millisecond pulsar PSR J0437$-$4715 \citep{2021ApJ...908..158S}, the magnetar XTE J1810$-$197 \citep{2024RMxAC..56...85A}, and seven pulsars that have previously exhibited glitches: PSR J0742$-$2822, PSR J0835$-$4510 \citep[a.k.a. the Vela pulsar;][]{2022MNRAS.509.5790L}, PSR J1048$-$5832, PSR J1644$-$4559, PSR J1721$-$3532, PSR J1731$-$4744, and PSR J1740$-$3015. Previous results from this campaign include: i) the confirmation of a glitch in PSR J0742$-$2822 \citep{2022ATel15638....1Z}; ii) the confirmation of the 2019 glitch in the Vela pulsar \citep{2019ATel12482....1L} and the announcement of its 2021 glitch \citep{2021ATel14806....1S, 2024RMxAC..56..161Z}, which we observed only one hour after it occurred; iii) the report of a glitch in PSR J1740$-$3015 that occurred in late 2022 \citep{2022ATel15838....1Z}; iv) the report of two small glitches in PSR J1048$-$5832 \citep{2023MNRAS.521.4504Z}, later confirmed by \cite{2023arXiv231204305L}; and v) the detection of the largest glitch reported to date in PSR J1048$-$5832 \citep{2024ATel16580....1Z}. Concerning the remaining pulsars (PSR J1644$-$4559, PSR J1721$-$3532 and PSR J1731$-$4744), none of them exhibited large glitches since the beginning of our monitoring campaign.
    
    Most giant glitches are characterized by a positive jump in frequency ($\Delta \nu > 0$) and a negative jump in the spin-down ($\Delta \dot \nu < 0$). However, smaller events can present every combination of signs \citep{2021A&A...647A..25E}, and they may actually be small glitches, part of the timing noise or some other phenomenon. Unfortunately, the statistics of these events are poor, as they are difficult to detect with low cadence monitoring. 
    In this work, we present a timing dataset of six pulsars in Sect.~\ref{sec:Pugliese} observed from the IAR with up to daily cadence. In Sect.~\ref{sec:method} we explain the algorithm we developed to perform a systematic search for small discrete events with all the possible combinations of signs in $\Delta \nu$ and $\Delta \dot \nu$ in the timing data of the pulsars. We present the results in Sect.~\ref{sec:results}, where we show the detected events and classify 3 of them as glitches after a more thorough analysis. In addition, we explore in Sect.~\ref{sec:results} the timing data of pulsars in which we already reported giant glitches in order to look for new recovery terms. We conclude with the discussion of the results in Sect. \ref{sec:discussion} and with some final remarks in Sect.~\ref{sec:conclusions}.

%
\section{Pulsar Glitch Monitoring Program at IAR}\label{sec:Pugliese}
%

\subsection{Observations}\label{sec:obs}

The IAR observatory, located near the city of La Plata, Argentina, has two 30-m single-dish antennas, “Carlos M. Varsavsky” (A1) and “Esteban Bajaja” (A2). 
A1 is configured to observe at a central frequency of 1400~MHz with a bandwidth of 112~MHz and one circular polarization, while A2 observes at a central frequency of 1428~MHz with a bandwidth of 56~MHz and two circular polarizations\footnote{We note that additional observations after April 2024 have an improved bandwidth of 400~MHz, but these are not part of this legacy dataset.}.

Since 2019, the PuMA collaboration has been using both antennas to observe a set of eight pulsars and one magnetar with up to daily cadence\footnote{\url{https://pugliese.iar.unlp.edu.ar/}}. In this work, we exclude the millisecond pulsar PSR~J0437$-$4715, which has an extremely stable rotation, the pulsar PSR~J1721$-$3532 and the magnetar XTE~J1810$-$197, as the timing accuracy achieved for them is only suitable for the detection of giant glitches (Araujo Furlan, in prep.). The number of observations collected for this work and their typical durations are shown in Table~\ref{tab: obs}. This dataset includes all available observations up to January 5th, 2024.

\begin{table*}[htbp]
    \centering
    \caption[]{Observations analysed in this work. }
    \label{tab: obs}
    \centering
    \begin{tabular}{cccccccccccc}
        \hline
        \noalign{\smallskip}
        Pulsar & \multicolumn{3}{c}{\# of observations} & Data span & $\langle{t_\mathrm{obs}\rangle}$ & $\langle{S/N\rangle}$ & $\langle{\sigma_\mathrm{TOA}\rangle}$ & $P$ & $\dot{P}$ & Age & Previous glitches\\
        & A1 & A2 & total & (MJD) & (min) & & ($\mu$s) & (s) & ($10^{-14}\, \mathrm{s}\, \mathrm{s^{-1}}$) & (kyr)\\
        \noalign{\smallskip}
        \hline
        \noalign{\smallskip}
        J0742$-$2822 & 345 & 408 & 753 & 58832--60149 & 27 & 6.6 & 190 & 0.167 & 1.68 & 157 & 9 \\
        J0835$-$4510 & 312 & 579 & 891 & 59264--60278 & 216 & 1024 & 7.1 & 0.089 & 12.5 & 11 & 22 \\
        J1048$-$5832 & 241 & 171 & 412 & 59031--60294 & 96 & 5.7 & 181 & 0.123 & 9.61 & 20 & 9 \\
        J1644$-$4559 & 229 & 403 & 632 & 58979--60294 & 35 & 7.3 & 35 & 0.455 & 2.01 & 359 & 4 \\
        J1731$-$4744 & 217 & 23 & 240 & 58751--60294 & 96 & 4.6 & 303 & 0.830 & 16.3 & 80 & 6 \\
        J1740$-$3015 & 0 & 316 & 316 & 59330--60314 & 114 & 8.4 & 286 & 0.606 & 46.6 & 20 & 38 \\
        \noalign{\smallskip}
        \hline
    \end{tabular}
    \tablefoot{
$\langle{t_\mathrm{obs}\rangle}$ is the typical observing time for each pulsar, and $\langle{S/N\rangle}$ and $\langle{\sigma_\mathrm{TOA}\rangle}$ are their typical S/N and TOA uncertainty, respectively. The number of previous glitches was extracted from \cite{2022MNRAS.510.4049B}.
}
\end{table*}

For all observations, we used the \textsc{PRESTO} package \citep{2003ApJ...589..911R, 2011ascl.soft07017R} to remove radio-frequency interferences (RFIs) with the task \texttt{rfifind} and to fold the observations with the task \texttt{prepfold}. The Times of Arrival (TOAs) were then calculated using the Fourier phase gradient-matching template fitting method described by \citep{1992RSPTA.341..117T}, implemented in the \texttt{pat} package in \texttt{psrchive} \citep{2004PASA...21..302H}. Given their similarities, we used the same template for observations of both antennas. We constructed this template by employing a smoothing wavelet technique (\texttt{psrsmooth} package in \texttt{psrchive}) on the pulse profile of a high signal-to-noise observation that was not part of the subsequent timing analysis.

\subsection{Pulsar timing}

Pulsar rotation is tracked by monitoring the TOAs of the pulses, while  a timing model is developed to predict the expected TOAs. 

In the timing model, a Taylor expansion is used to model the temporal evolution of the pulsar phase,
\begin{equation}\label{eq:timing-model}
    \phi(t)=\phi+\nu(t-t_0)+\frac{1}{2}\dot{\nu}(t-t_0)^2+\frac{1}{6}\Ddot{\nu}(t-t_0)^3,
\end{equation}
where $\nu$, $\dot\nu$ and $\ddot\nu$ are the rotation frequency of the pulsar, and its first and second derivatives, respectively, and $t_0$ is the reference epoch. Diverse physical phenomena can be studied through this technique, including the internal structure of pulsars that gives rise to glitches \citep{2022RPPh...85l6901A}.

During a glitch, there is a sudden jump in the pulsar rotation frequency. The additional phase induced by a glitch can be described as \citep{2013MNRAS.429..688Y}:
\begin{multline}\label{eq:glitch-model}
    \phi_\mathrm{g}(t) = \Delta \phi + \Delta \nu_\mathrm{p} (t-t_\mathrm{g}) + \frac{1}{2} \Delta \dot{\nu}_\mathrm{p} (t-t_\mathrm{g})^2 + \\ 
    \frac{1}{6} \Delta \Ddot{\nu}(t-t_\mathrm{g})^3 + \left[1-\exp{\left(-\frac{t-t_\mathrm{g}}{\tau_\mathrm{d}}\right)} \right]\Delta \nu_\mathrm{d} \, \tau_\mathrm{d},
\end{multline}

\noindent where $t_\mathrm{g}$ denotes the glitch epoch, $\Delta \phi$ counteracts the uncertainty on $t_\mathrm{g}$, and $\Delta \nu_\mathrm{p}$, $\Delta \dot{\nu}_\mathrm{p}$ and $\Delta \Ddot{\nu}$, are the respective permanent shifts in $\nu$, $\dot\nu$ and $\Ddot{\nu}$ relative to the pre-glitch solution. Additionally, $\Delta \nu_\mathrm{d}$ stands for the temporary increase in frequency that recovers exponentially over a duration of $\tau_\mathrm{d}$. Considering the instantaneous frequency shift as $\Delta \nu_\mathrm{g} = \Delta \nu_\mathrm{p} + \Delta \nu_\mathrm{d}$ and the instantaneous shift in the frequency derivative as $\Delta \dot{\nu}_\mathrm{g} = \Delta \dot{\nu}_\mathrm{p} - \frac{\Delta \nu_\mathrm{d}}{\tau_\mathrm{d}}$, one can compute the recovery coefficient, $Q$, which relates the temporary and permanent frequency shifts as $Q=\Delta \nu_\mathrm{d} / \Delta \nu_\mathrm{g}$, and the total instantaneous jumps in $\nu$ and $\dot\nu$ as $\Delta \nu_\mathrm{g} / \nu$ and $\Delta \dot\nu_\mathrm{g} / \dot\nu$ respectively. 

Initially, we took the parameters for the timing models from the ATNF pulsar catalogue \citep{2005AJ....129.1993M}, and then we updated them by fitting our data to the timing model (Eq. \ref{eq:timing-model}) with the \textsc{Tempo2} \citep{2006MNRAS.369..655H} software package. Pulsars that present abundant timing noise or have undergone glitches could be difficult to solve coherently. However, the high-cadence of our observations allows phase-connected timing solutions without adding any additional phase jump.

\subsection{Data release}

\begin{figure}[h!]
    \includegraphics[width=\linewidth]{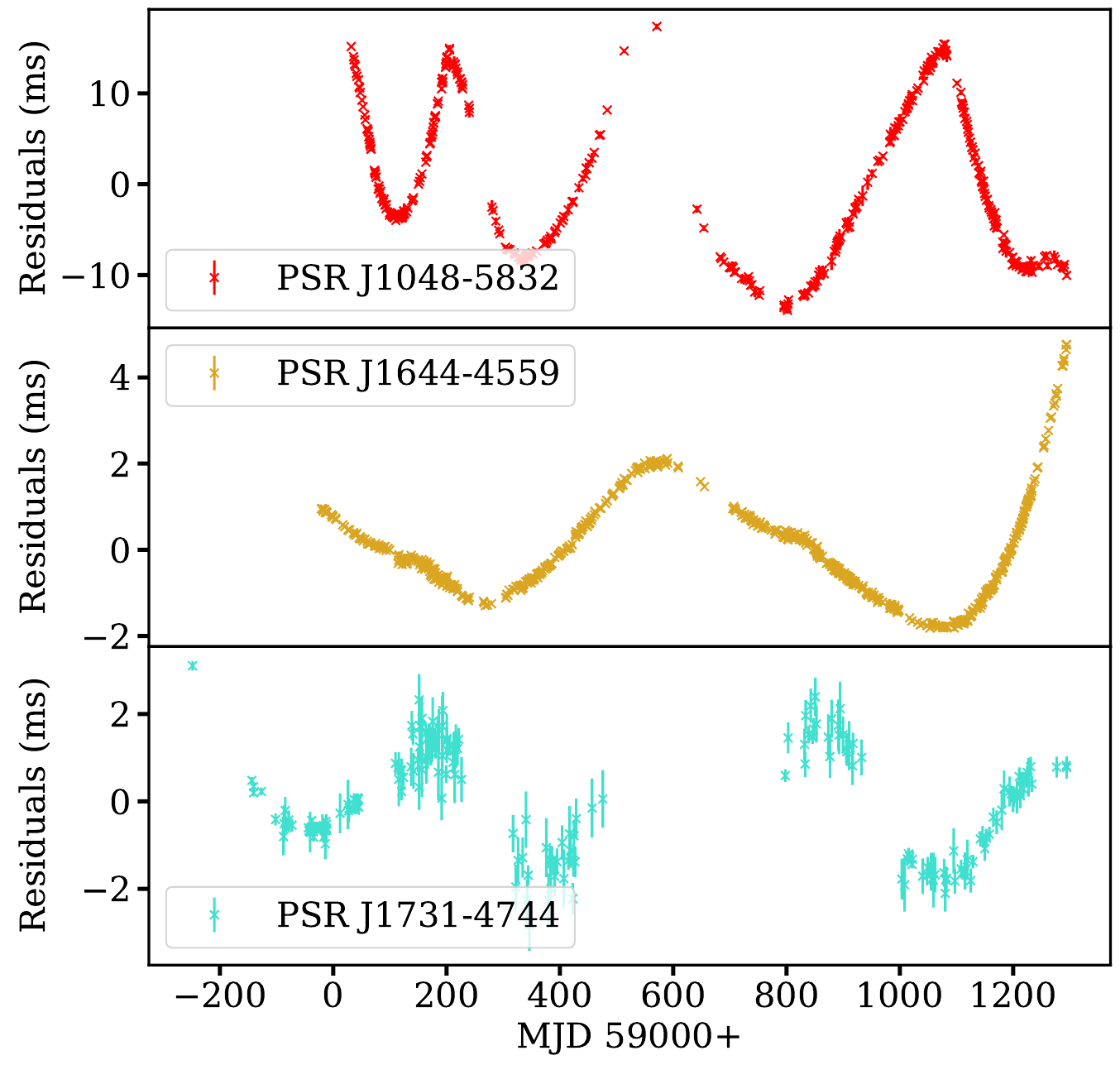}
    \caption{Residuals for pulsars that did not have a giant glitch during our campaign.}
    \label{fig: residuals1}
\end{figure}

\begin{figure}[h!]
    \includegraphics[width=\linewidth]{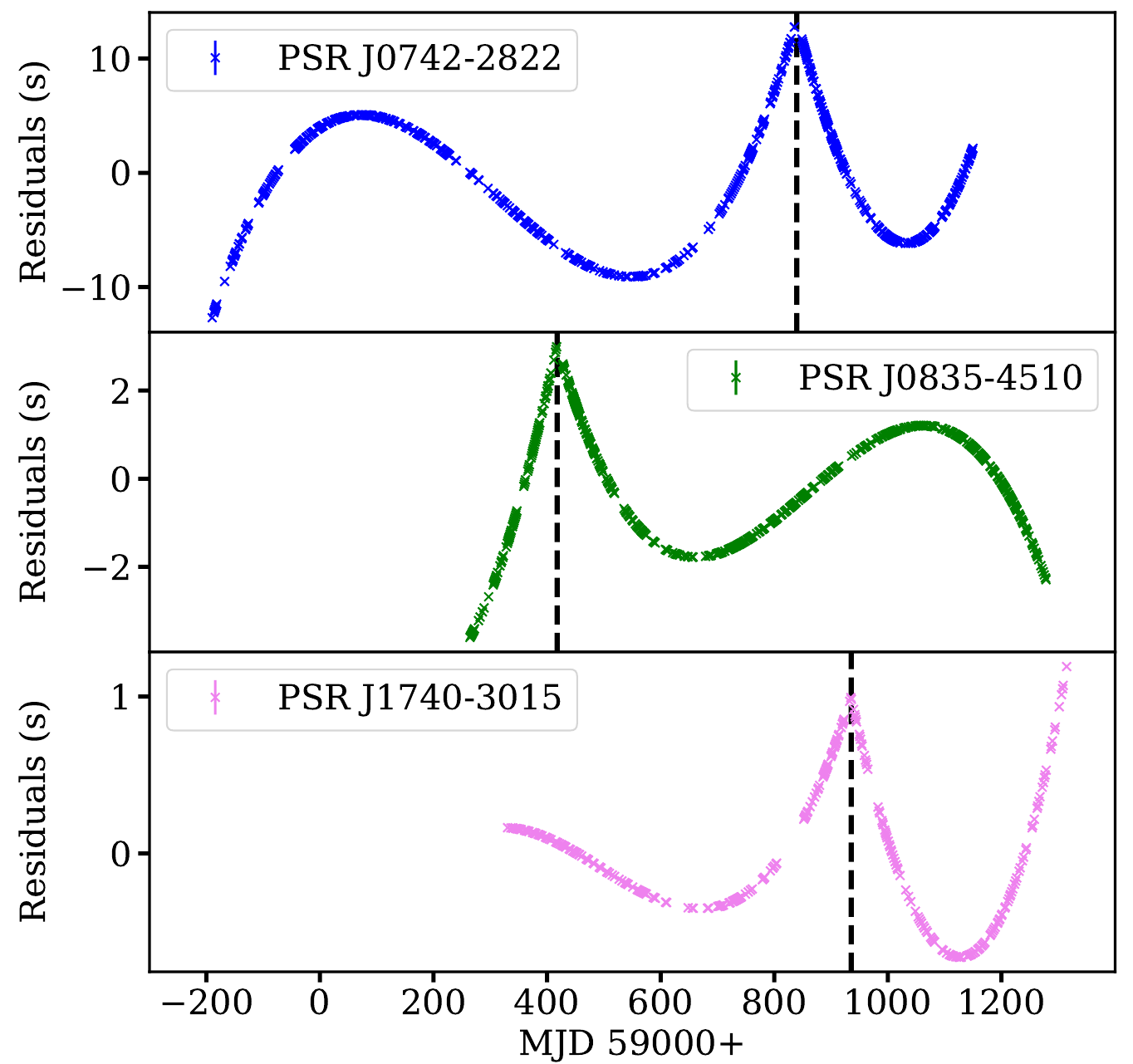}
    \caption{Residuals for pulsars that experienced a giant glitch during our campaign.}
    \label{fig: residuals2}
\end{figure}

We used our TOAs to update the ephemeris for each pulsar. The TOAs and up-to-date timing solutions can be found at our online repository\footnote{\url{https://github.com/PuMA-Coll/Timing_irregularities}}. The residuals for pulsars that have not undergone a giant glitch beneath our data span before January 5th 2024 (MJD 60314) are shown in Fig. \ref{fig: residuals1}. In Fig. \ref{fig: residuals2} we show the residuals of pulsars with giant glitches beneath our data span.

To account for jitter or systematic errors we included the parameters EFAC and EQUAD, which model white noise modifying each TOA uncertainty as
\begin{equation}
    \label{eq:sigma_toa}
    \sigma_{\rm TOA} = \sqrt{{\rm EQUAD}^2 + {\rm EFAC} \times \sigma_i^2},
\end{equation}
where $\sigma_i$ is the TOA uncertainty for each observation derived from the cross-correlation between the template profile and the folded observation. EFAC captures the impact of unaccounted-for instrumental effects and imperfect estimations of TOA uncertainties, while EQUAD addresses any additional sources of time-independent uncertainties, such as pulse jitter.

\begin{figure}[h!]
    \includegraphics[width=\linewidth]{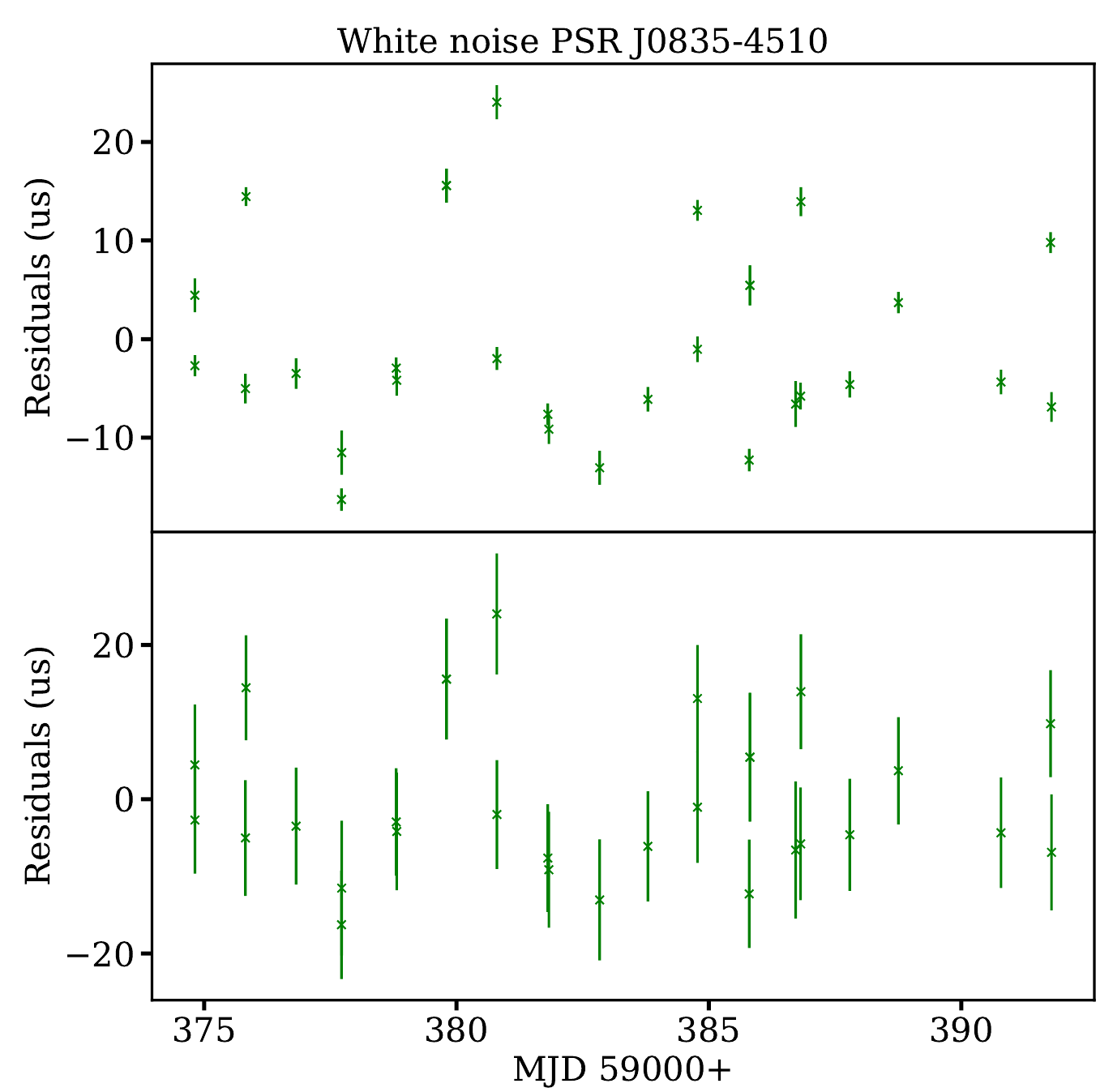}
    \caption{Process to adjust the TOAs uncertainties. \textit{Top:} We keep a short data span with flat residuals. \textit{Bottom:} We add the parameters EFAC and EQUAD (Eq.~\ref{eq:sigma_toa}) calculated with \texttt{TempoNest}.}
    \label{fig: white_noise}
\end{figure}

To obtain EFAC and EQUAD for each pulsar we fitted $\nu$ and $\dot\nu$ in a data span short enough to obtain flat residuals, as shown in Fig. \ref{fig: white_noise}. Then we used \texttt{TempoNest} \citep{2014MNRAS.437.3004L} to obtain the white noise of the pulsar. 

%
\section{Timing irregularities detection}\label{sec:method}
%

\subsection{Search sensitivity}

\begin{figure}[h!]
    \includegraphics[width=\linewidth]{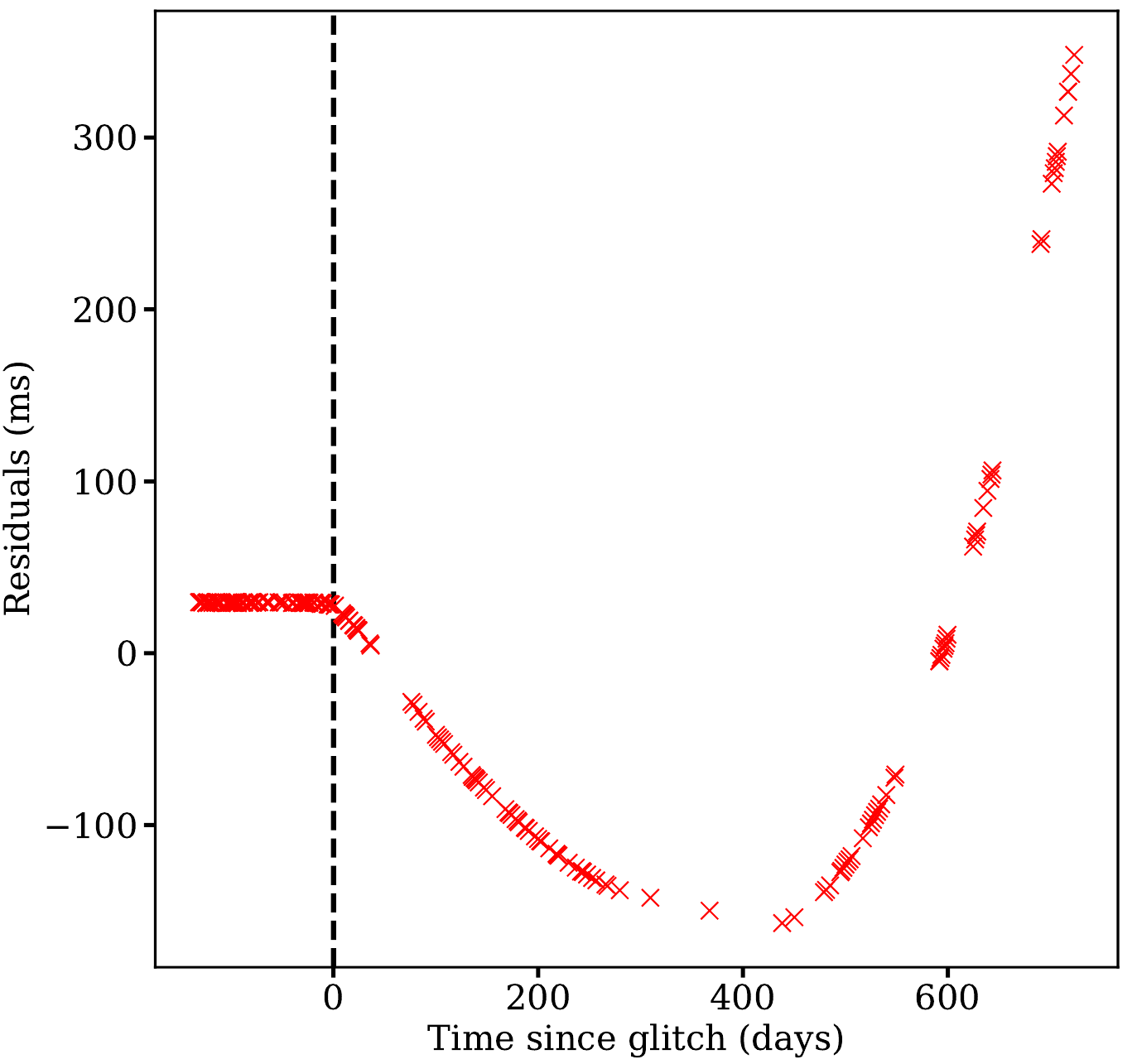}
    \caption{Signature of a glitch with $\Delta \nu > 0$ and $\Delta \dot \nu <0$ in the residuals. The residuals correspond to the glitch we reported on MJD 59204 in PSR J1048$-$5832 in \cite{2023arXiv231208188Z}.}
    \label{fig: glitch-example}
\end{figure}

We estimated the detection limits for our results following \cite{2014MNRAS.440.2755E}. When a positive frequency jump (i.e., glitch) occurs, the residuals start to diverge linearly towards negative values. Then, if there is also a negative change in the spin-down rate, the residuals gradually rise towards positive values with a parabolic behaviour. This behaviour is depicted in Figure~\ref{fig: glitch-example}. 

To ensure the detection of a glitch, the cadence of the observations must be high enough to have at least one TOA before the residuals become positive again. Additionally, the most negative residual value must exceed the TOAs uncertainties. Consequently, we define our detection boundaries for events with $\Delta \nu > 0$ and $\Delta \dot \nu <0$ using the following equation, which accounts for both conditions:
\begin{equation}\label{eq: boundaries}
\Delta \nu_\mathrm{lim} = \mathrm{max} \left( \, \Delta T \lvert\Delta \dot \nu \rvert /2 \,,\, \sqrt{2 \lvert \Delta \dot \nu \rvert \sigma_{\rm TOA}} \,
\right).
\end{equation}
This criterion is also applicable to events with $\Delta \nu < 0$ and $\Delta \dot \nu >0$. We assume a daily monitoring cadence ($\Delta T=1$~d) in the analysis of Sect.~\ref{sec:results}.

On the other hand, for events where $\Delta \nu $ and $\Delta \dot \nu $ have the same sign, the only criterion is that the TOAs uncertainties  should be smaller than the residuals. 
Considering these criteria and Equation 4, we can define our sensitivity in detecting events with any combination of signs. 
In our observations, we find that the detection limit for cases with different signs in $\Delta \nu$ and $\Delta \dot \nu$ is primarily determined by the uncertainties associated with the TOAs, rather than the observation cadence.

\subsection{Method}\label{sec: method}

During a glitch, there is a sudden positive jump in $\nu$. In this case, residuals after the glitch diverge linearly to negative values, and they recover quadratically if there is also a discontinous change in $\dot \nu$ of the opposite sign. In the case of an anti-glitch, the jump in $\nu$ is negative and the change in $\dot \nu$ is positive, so residuals diverge linearly to positive values and then recover quadratically. Hence, one method for glitch detection is to set $t_0$ at the suspected glitch epoch and fit $\nu$ and $\dot \nu$ to the data before and after $t_0$; we denote the values fitted to the data after $t_0$ as $\nu'$ and $\dot\nu'$. The comparison between $\nu$ and $\nu'$ can reveal the presence of a glitch.

\begin{figure}[h!]
\centering
    \includegraphics[width=0.9\linewidth]{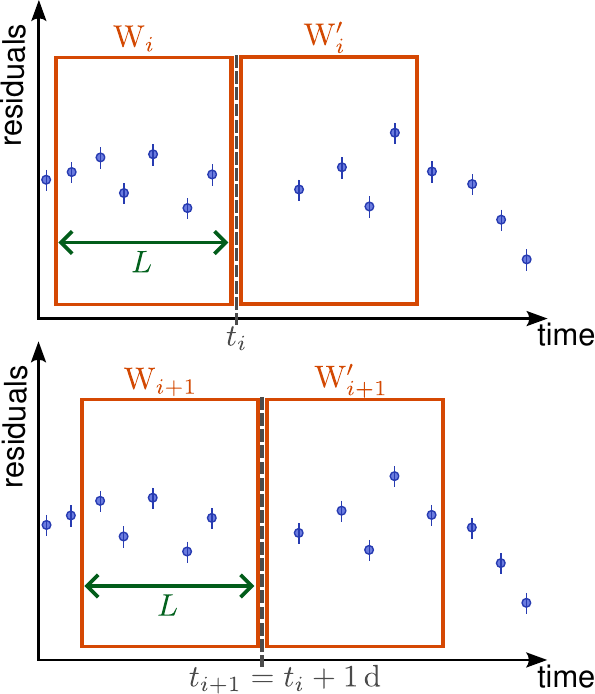}
    \caption{Schematic representation of steps 1 and 2 of the algorithm described in Sect.~\ref{sec: method}. Example with $N_\mathrm{min}=5$. \textit{Top:} $\nu$ and $\dot \nu$ are not fitted in windows $W_i$ and $W'_i$ because there are only four TOAs inside $W'_i$. \textit{Bottom:} $\nu$ and $\dot \nu$ are fitted in windows $W_{i+1}$ and $W'_{i+1}$ because there are at least five TOAs inside each of them.}
    \label{fig: algorithm}
\end{figure}

We developed an algorithm based on \textsc{PINT} \citep{2021ApJ...911...45L, 2019ascl.soft02007L} to look for discrete jumps in frequency in our pulsar timing data\footnote{\url{https://github.com/PuMA-Coll/Timing_irregularities}}. The algorithm works as follows (see Fig.~\ref{fig: algorithm} for a schematic representation):
\begin{enumerate}
    \item Restrict the timing data to a small data span of $2L$ days and set the parameter $t_0$ (see Eq.~\ref{eq:timing-model}) at the middle of that data span. That is, for a data span [MJD X, MJD X+$2L$], define $t_i =$~MJD~X+$L$ and set $t_0=t_i$.
    \item If the number of TOAs in [MJD X, MJD X+$L$] or [MJD X+$L$, MJD X+$2L$] is lower than $N_\mathrm{min}$, go to step 6.
    \item Fit $\nu$ and $\dot{\nu}$ (with their uncertainties $\sigma_{\nu}$ and $\sigma_{\dot{\nu}}$) using only TOAs within [MJD X, MJD X+$L$].
    \item Fit $\nu'$ and $\dot{\nu}'$ (with their uncertainties $\sigma_{\nu'}$ and $\sigma_{\dot{\nu'}}$) using only TOAs within [MJD X+$L$, MJD X+$2L$], keeping fixed the value of $t_0$.
    \item Calculate $ \Delta\nu = \nu'-\nu$ and $\Delta \dot \nu = \dot{\nu}'-\dot{\nu}$. If $\lvert \Delta\nu \rvert > \sigma_{\Delta\nu}$, save the values $ \Delta\nu$ and $ \Delta\dot\nu$.
    \item Repeat from 1 to 5 with [MJD X+1, MJD X+$2L$+1] and $t_0=t_i+1~\mathrm{d}$.
    \item Finally, if two detections are closer than $L$ days, keep only the one with the highest value of $\lvert \Delta\nu \rvert$.
\end{enumerate}
Here $N_\mathrm{min}$ and $L$ are user-specified parameters. We tested different values and concluded that $L=24$ and $N_\mathrm{min}=6$ give a good compromise between having good statistics in each window and not averaging out small events. The minimum size of $\Delta \nu$ that we can detect is defined by Eq.~\ref{eq: boundaries}. We note that we keep $\Ddot\nu$ fixed at the value fitted with the whole data span.

%
\section{Results}\label{sec:results}
%

\begin{figure*}[h!]
    \centering
    \includegraphics[height=0.3\textheight,width=0.46\linewidth]{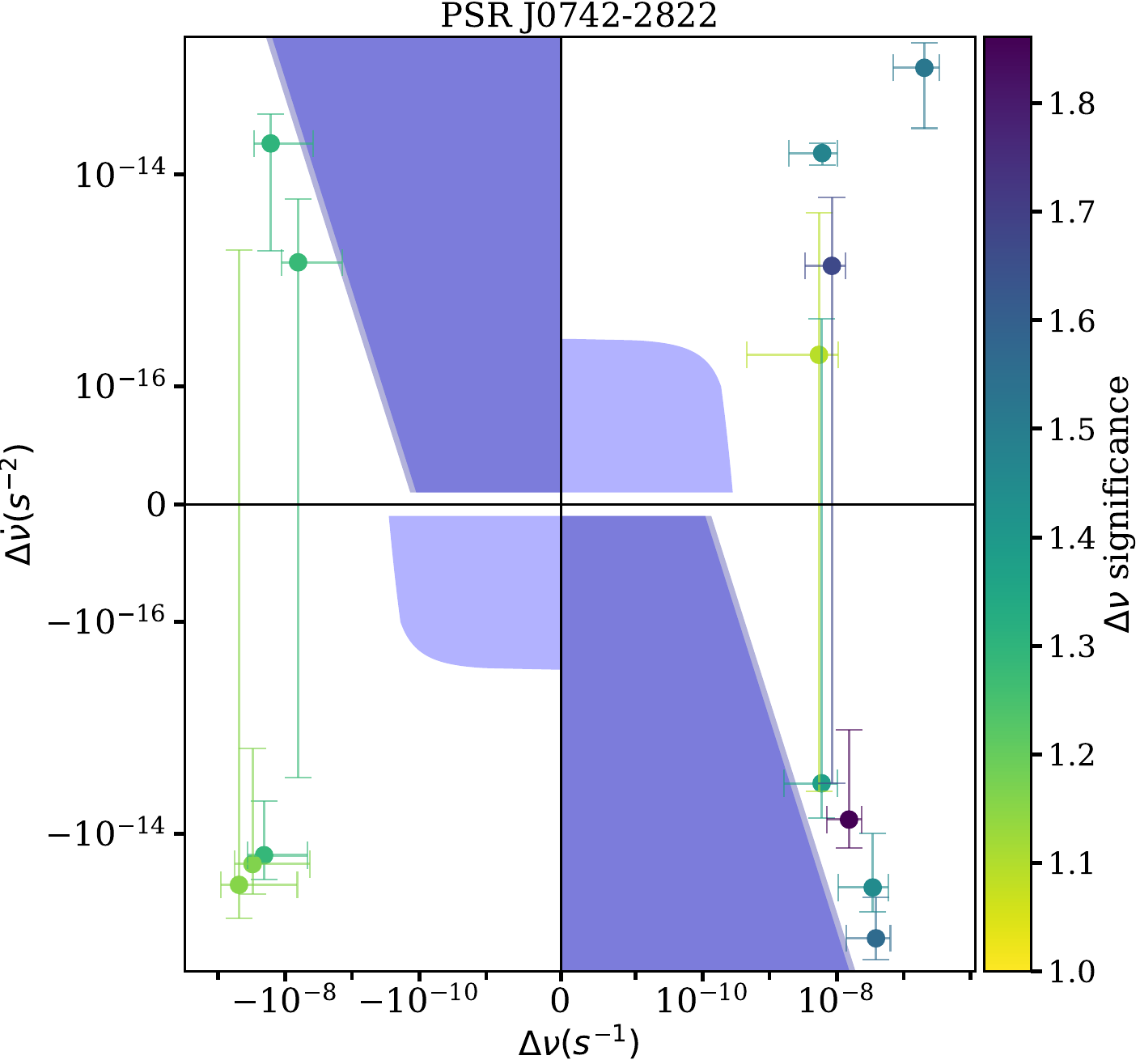} \hfill
    \includegraphics[height=0.3\textheight,width=0.46\linewidth]{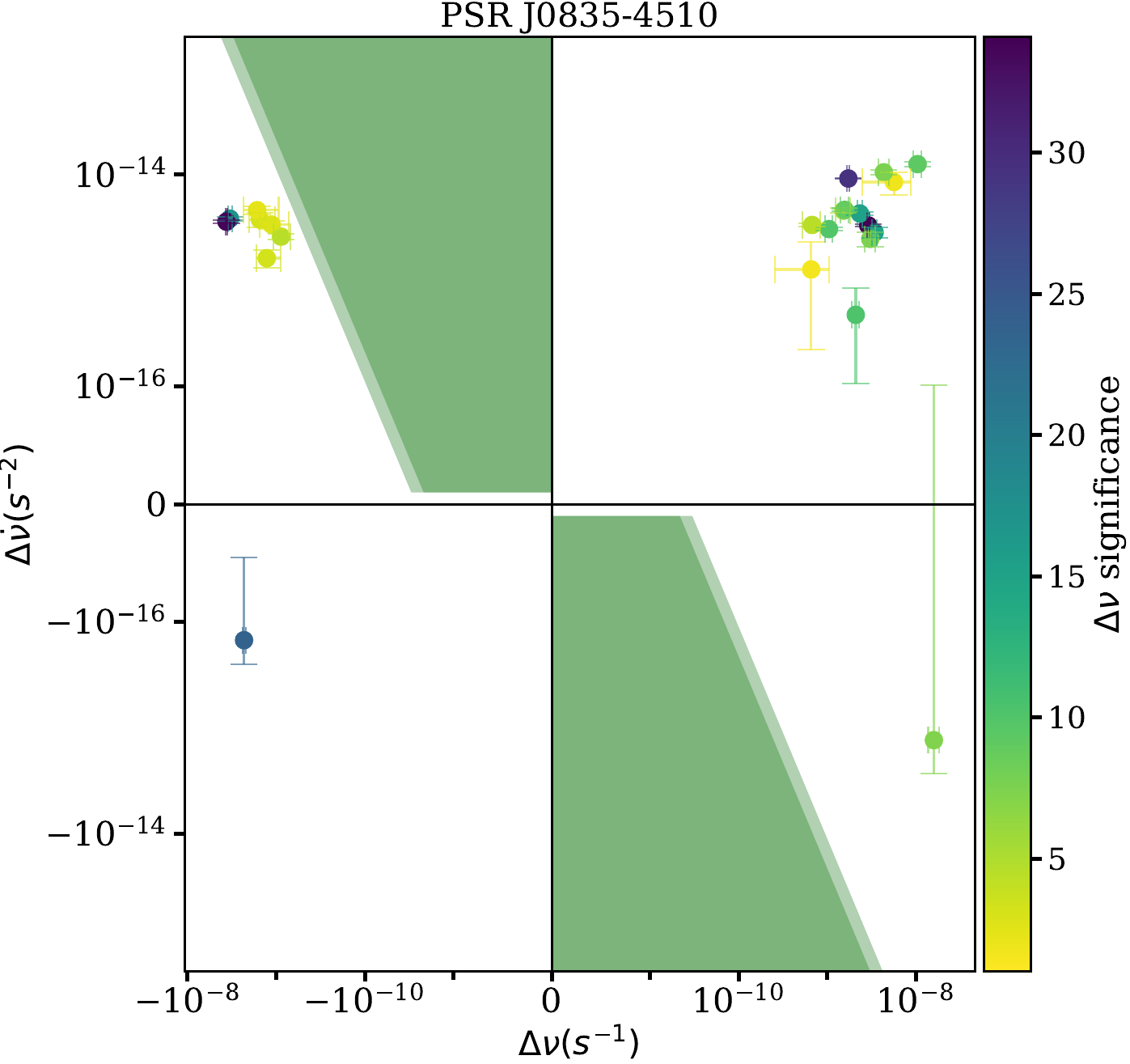}
    \\
    \includegraphics[height=0.3\textheight, width=0.46\linewidth]{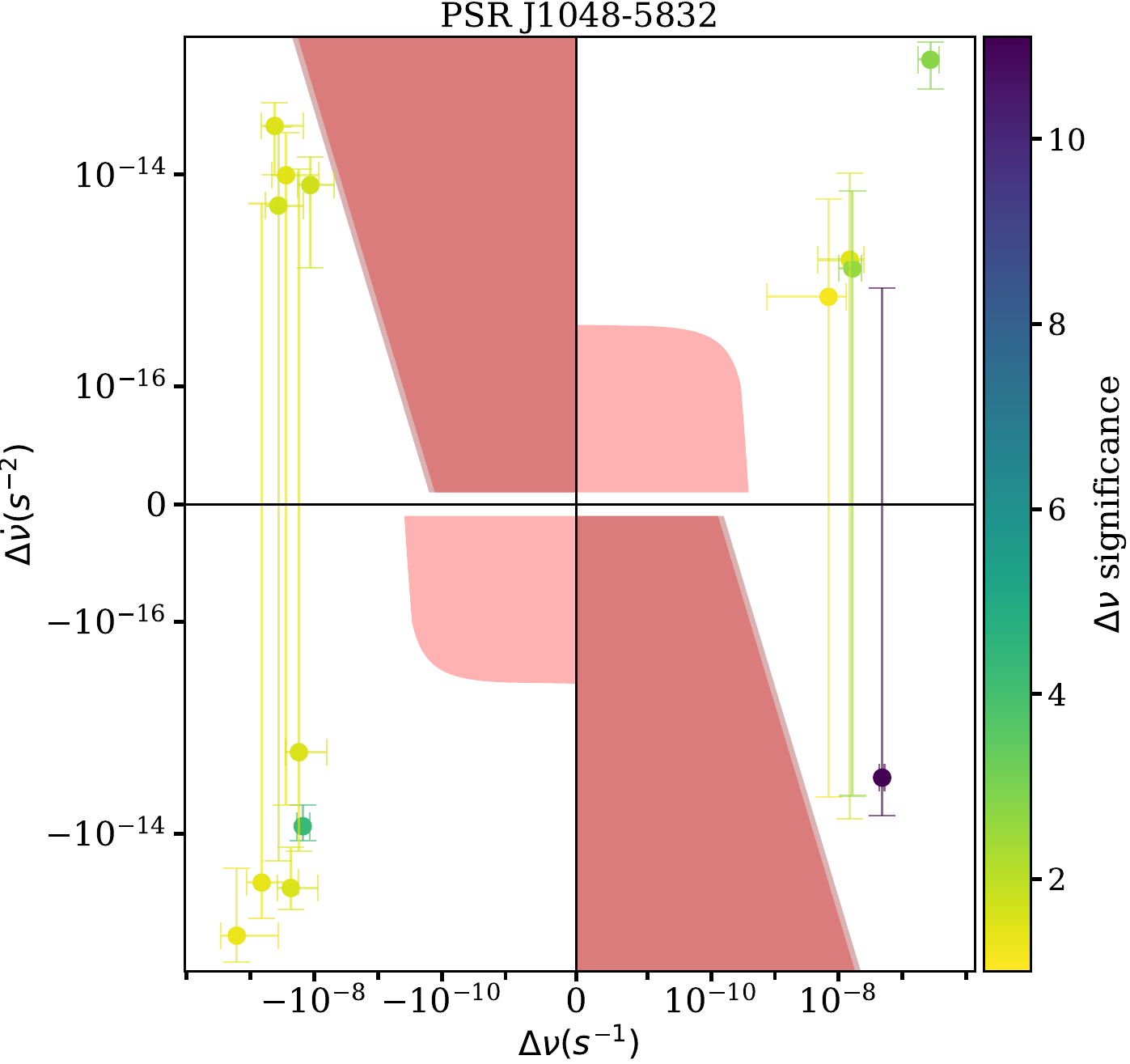} \hfill
    \includegraphics[height=0.3\textheight,width=0.46\linewidth]{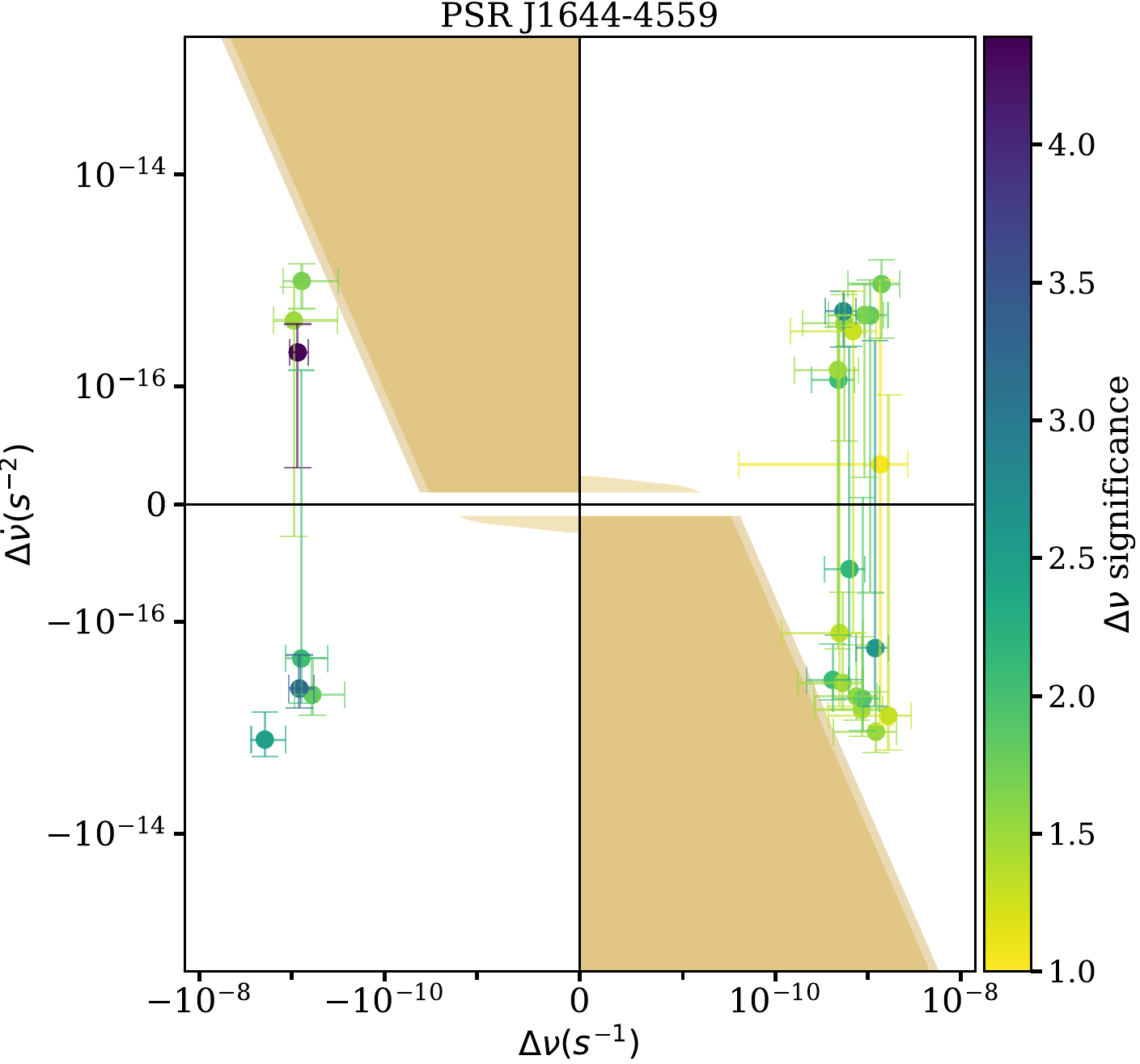}
    \\
    \includegraphics[height=0.3\textheight,width=0.46\linewidth]{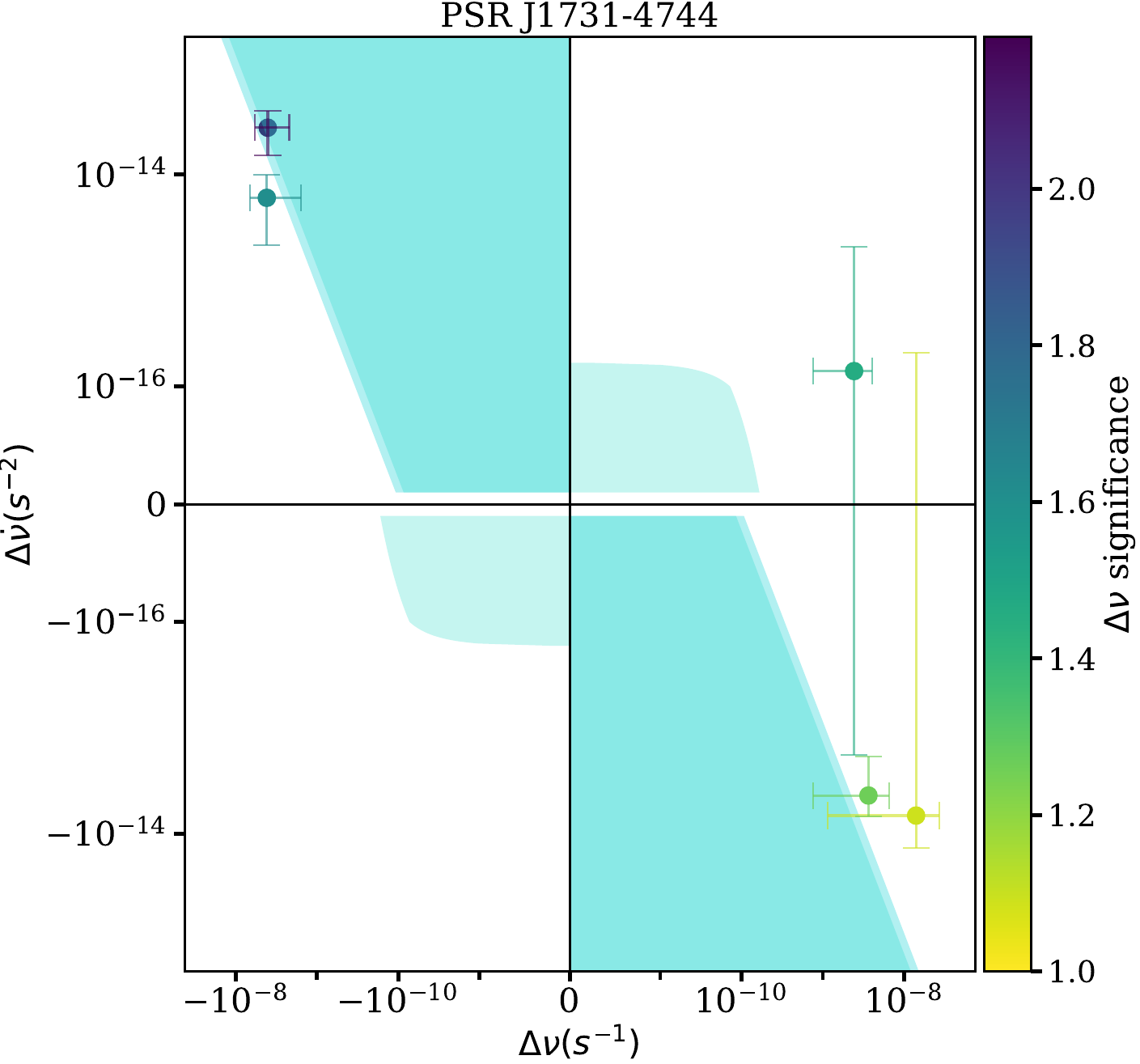} \hfill
    \includegraphics[height=0.3\textheight,width=0.46\linewidth]{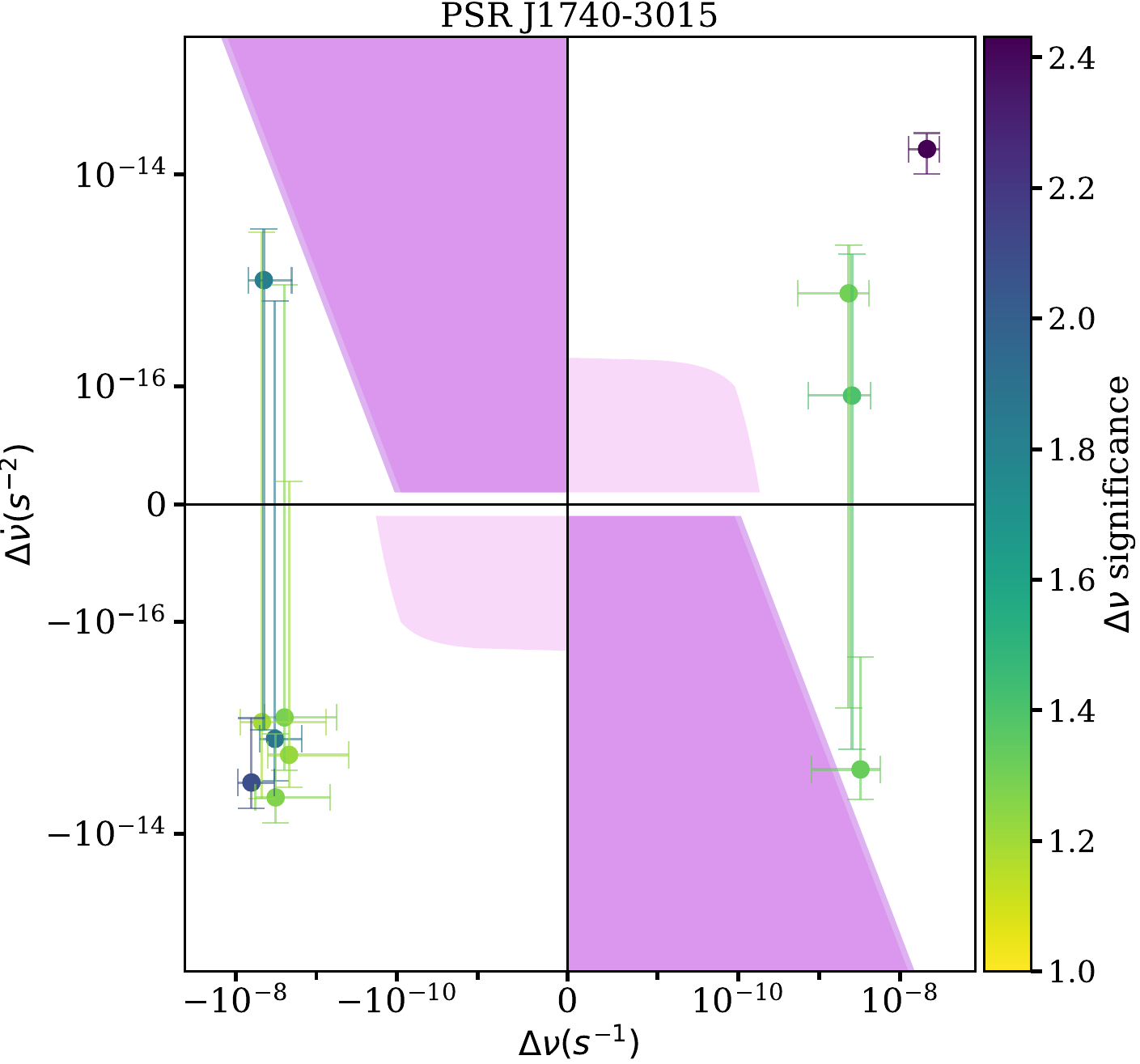}
    \caption{Irregularities detected by our algorithm in the six pulsars. The colorbar indicates the significance of the detection. The two shaded regions correspond to detection limits following Eq.~\ref{eq: boundaries} calculated with $\sigma_{\rm TOA}$ and $\sigma_{\rm TOA} + \Delta \sigma_{\rm TOA}$ respectively.}
    \label{fig: result}
\end{figure*}

In Fig.~\ref{fig: result} we show the detections associated with the six glitching pulsars. We excluded the detections of giant glitches in PSR J0742$-$2822, PSR J0835$-$4510 and PSR J1740$-$3015, in order to better appreciate the distribution of smaller events detected. All of them were detected in quadrant IV as it is usual for giant glitches.

To simplify the discussion, we will refer to quadrant I to the top right quadrant in each plot ($\Delta\nu$ ,$\Delta\dot{\nu}$ >0), quadrant II to the top left ($\Delta\nu$<0, $\Delta\dot{\nu}$>0), quadrant III to the bottom left ($\Delta\nu$ , $\Delta\dot{\nu}$<0) and quadrant IV to the bottom right ($\Delta\nu$>0, $\Delta\dot{\nu}$<0). Besides analyzing size distribution of events, we also studied in detail each of them. To change the status of an event to a glitch, we used the \texttt{glitch} plug-in in \texttt{tempo2} \citep{2006MNRAS.369..655H} to subdivide the observations into blocks spanning between 6 and 10 days, ensuring that each block contains at least 6 TOAs and then fit $\nu_0$ and $\dot\nu_0$ in each of these blocks. We classify an event as a glitch when we observe a step-like change in $\nu_0$ and the residuals can not be adequately fitted using a second order polynomial model, but instead require fitting with a step change function. If we do not detect a step-like change in $\nu_0$, we change the event status to irregularity.

To obtain detection limits, we used Eq.~\ref{eq: boundaries} considering the average of all $\sigma_{\rm TOA}$ for each pulsar and its standard deviation ($\Delta \sigma_{\rm TOA}$). We show such detection limits as shaded regions in Fig.~\ref{fig: result}, where the smaller regions where calculated with $\sigma_{\rm TOA}$ and the larger regions were calculated assuming $\sigma_{\rm TOA} + \Delta \sigma_{\rm TOA}$.


\subsection{PSR J0742--2822}
We detected 13 events: four events in quadrant I, two in quadrant II, three in quadrant III and four in quadrant IV. Despite the significance of some events with $\Delta\nu>0$ and $\Delta\dot{\nu}<0$, we did not find glitch signatures in the residuals for any of them.%
We note that eight irregularities have $\Delta \nu>0$ while only five have $\Delta \nu<0$. 
These jumps in the rotation frequency are accompanied by opposite-signed, same-signed or null alterations in $\Delta \dot\nu$. All events oscillate between $ 5\times10^{-9} <\lvert \Delta\nu (s^{-1}) \rvert < 2\times 10^{-7}$ and $ 10^{-16}<\lvert \Delta\dot\nu (s^{-2}) \rvert < 10^{-13}$.

\begin{figure}[h!]
    \includegraphics[width=\linewidth]{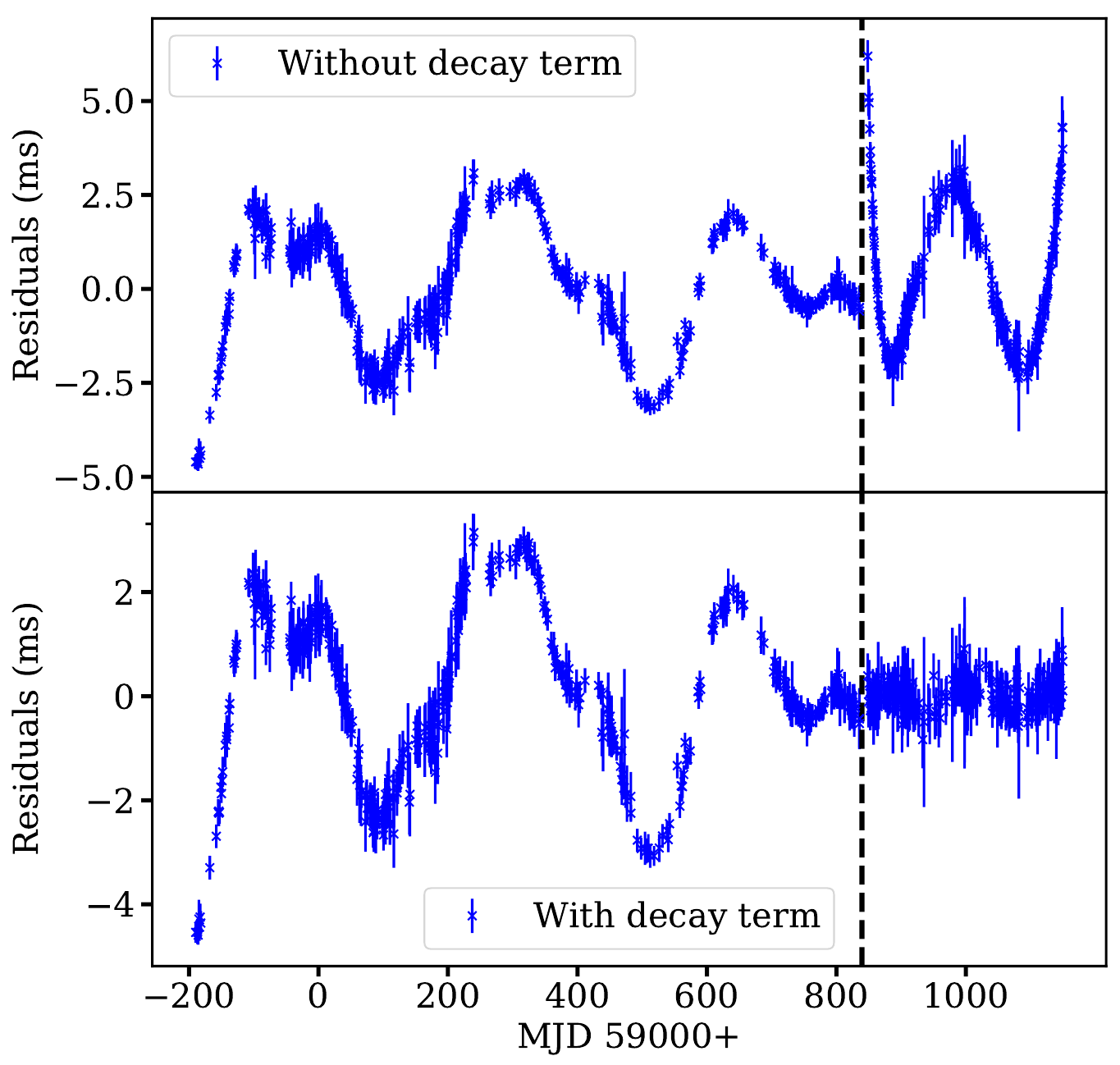}
    \caption{Detection of a new decay term in the PSR J0742$-$2822 glitch on MJD 59839.4(5). \textit{Top:} Residuals without considering any decay term. \textit{Bottom:} Residuals considering one decay term of $33.4(5)~\mathrm{d}$.}
    \label{fig: J07-glitch}
\end{figure}

Our algorithm also detected the last giant glitch in PSR J0742$-$2822, which was excluded from Fig. \ref{fig: result}. This glitch was initially reported by \cite{2022ATel15622....1S}, and then we reported our confirmation \citep{2022ATel15638....1Z, 2023MNRAS.521.4504Z} of a giant glitch of $\Delta\nu_g/\nu=4295(1)\times10^{-9}$ on MJD=59839.4(5). However, our post-glitch data span was not sufficient to detect any long-term exponential recovery. In this current work with a much longer post-glitch data span, we assumed $t_g=59839.4(5)$ as reported in \citep{2022MNRAS.510.4049B}\footnote{\url{http://www.jb.man.ac.uk/pulsar/glitches.html}} and we find one recovery term of $\tau = 33.4(5)~\mathrm{d}$ with a degree of recovery of $Q=0.446(1)\%$, shown in Fig.~\ref{fig: J07-glitch}. Also, using a combination of PINT \citep{2021ApJ...911...45L} and an MCMC-like code to find the solutions, we calculated the evidence for a model considering a change in $\Ddot \nu$ and the one without it. The Bayes factor for the model with $\Delta \Ddot \nu$ over the one without it yielded $1.2 \times10^{-8}$, which indicates that the model with $\Delta \Ddot \nu=0$ is strongly preferred. The updated parameters considering the new recovery term can be seen in Table \ref{tab:updated_parameters}.

Fig.~\ref{fig: J07-glitch} shows that, once the glitch model is subtracted, the residuals flatten after the glitch. This has two strong implications: the post-glitch behaviour is explained with only one decay component of $33.4(5)~\mathrm{d}$ and that the red noise component significantly diminishes after the glitch at least for the 300d covered by our observations.

\begin{figure}[h!]
    \includegraphics[width=\linewidth]{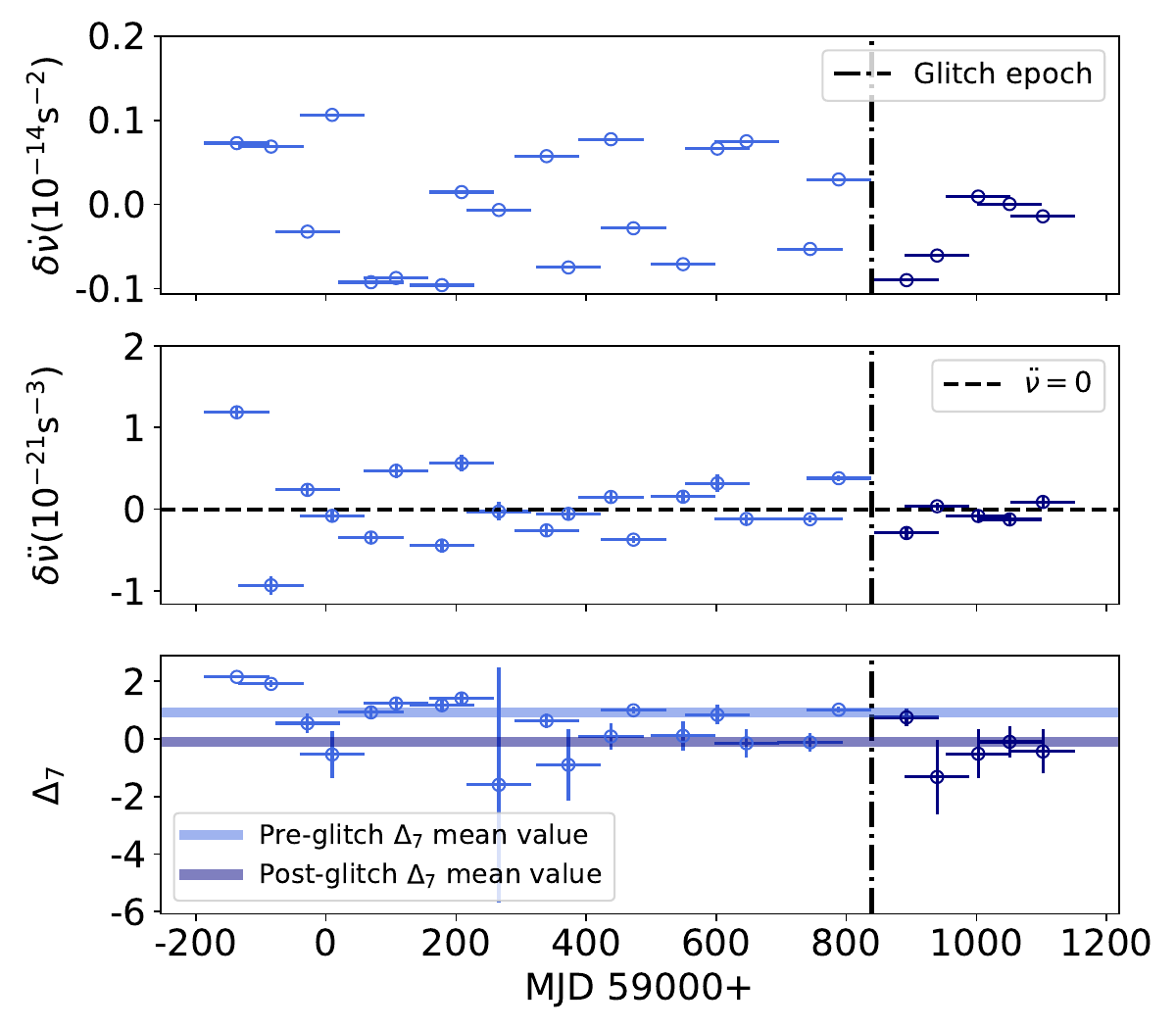}
    \caption{Timing parameters of J0742$-$2822 calculated in $100~\mathrm{d}$ windows overlapping by 50 TOAs between consecutive windows; the time window is represented with an horizontal errorbar. \textit{Top:} Evolution of $\dot\nu$ subtracting glitch model (Eq. \ref{eq:glitch-model}). \textit{Center:} Evolution of $\Ddot \nu$ subtracting glitch model. \textit{Bottom:} Evolution of $\Delta_7$ calculated from $\nu$ and $\Ddot{\nu}$ with the glitch model subtracted (Sec.~\ref{sec:discussion}). The horizontal lines corresponds to the mean pre-glitch value of $\Delta_7$ and the mean post-glitch value of $\Delta_7$.}
    \label{fig: J07-noise}
\end{figure}

 To quantify the change in red noise before and after the glitch, we explored the behaviour of $\dot \nu$ and $\Ddot \nu$, subtracting the glitch model from the data. Following \cite{1994ApJ...422..671A}, we also calculated the parameter $\Delta_n(t) = \log{\left[ \lvert\Ddot \nu\rvert t^3 / (6 \nu) \right]}$, where $t \simeq 10^{n}~\mathrm{s}$. This parameter quantifies the strength of the timing noise. We adopted $n=7$ and show the results in Fig.~\ref{fig: J07-noise}.

Fig.~\ref{fig: J07-noise} shows that the amplitude of the oscillations in $\dot \nu$ and $\Ddot \nu$ is larger pre-glitch than post-glitch. Also, $\Ddot \nu$ stabilizes around zero after the glitch, and pre-glitch mean value of $\Delta_7$ is above the post-glitch mean value of $\Delta_7$, which indicates timing noise is stronger before the glitch. 

\subsection{PSR J0835--4510}

PSR J0835$-$4510 (Vela pulsar) is the most studied pulsar of the southern hemisphere, given the periodicity of its giant glitches ($\sim 3~\mathrm{yr}$) . Our long observations of the bright Vela pulsar leads to events detected with a higher significance (up to 30$\sigma$), as seen in Fig.~\ref{fig: result}.

We detected 23 events. It is particularly interesting that, with the exception of two of them, all events are in the quadrants I or II, which means that most of the events have a positive $\Delta\dot\nu$. After each giant glitch, Vela shows a negative jump in $\dot\nu$, which recovers exponentially at the beginning, and then linearly until a value near the pre-glitch value of $\dot\nu$ is reached. At this point, another glitch is expected to occur \citep{2022MNRAS.511..425G}. It may be that this positive trend (in which $\ddot \nu>0$) is not perfectly smooth and there are instances in which $\dot\nu$ increases faster; thereby producing discrete events that are found by our method of detection.

In particular, all events lie in the range of $ 6\times 10^{-10}<\lvert \Delta\nu (s^{-1}) \rvert < 2 \times 10^{-8}$ and $10^{-16}<\lvert \Delta\dot\nu (s^{-2}) \rvert < 1.3 \times 10^{-14}$.

\begin{figure}[h!]
    \includegraphics[width=\linewidth]{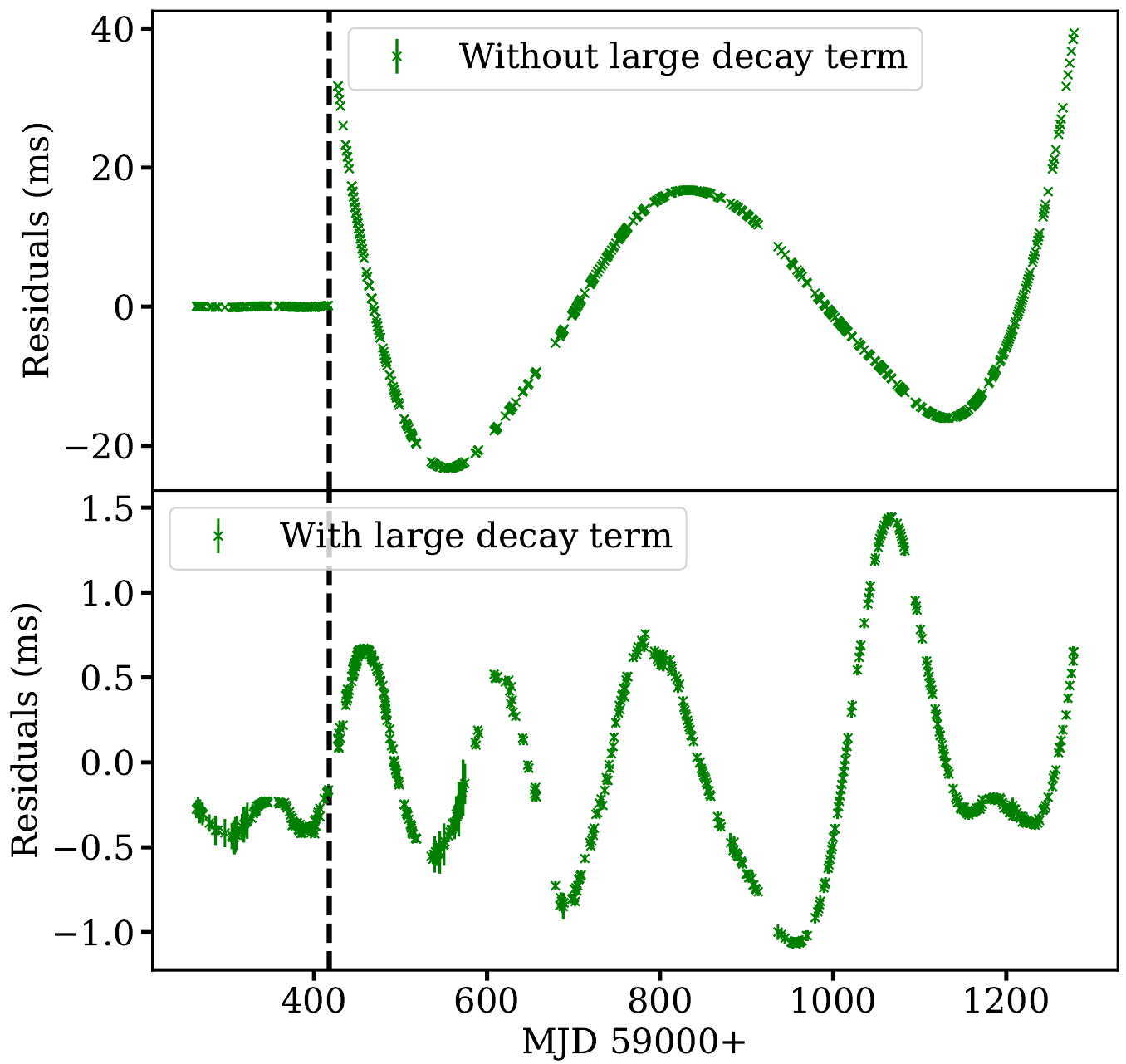}
    \caption{Detection of a new decay term in the 2021 Vela glitch. \textit{Top:} Residuals without considering long decay term. \textit{Bottom:} Residuals considering the long decay term of $535(8)~\mathrm{d}$. Note the much smaller residual scale. The signal that remains is the timing noise of the Vela pulsar.}
    \label{fig: J08-glitch}
\end{figure}

Our algorithm also detected the 2021 Vela giant glitch, which we excluded from Fig. \ref{fig: result}. We announced this glitch in \cite{2021ATel14806....1S} and, later on, we reported in \cite{2023MNRAS.521.4504Z} the presence of two short exponential decay components associated with that giant glitch: $\tau_1=6.400(2)~\mathrm{d}$ (with $Q_1=0.2(1)\%$) and $\tau_2=0.994(8)~\mathrm{d}$ (with $Q_2=0.7(1)\%$). Here
we report the detection of a third recovery term with a much longer time scale, $\tau_3 = 535(8)~\mathrm{d}$, and a significant degree of recovery of $Q_3=41(1) \%$. We show the corresponding residuals in Fig.~\ref{fig: J08-glitch}, and updated parameters for the glitch in Table \ref{tab:updated_parameters}.

Between our 23 detectiones mentioned above, we detected a small event around MJD 59479, which is 62 days after the giant glitch. This event was initially fitted to quadrant I by the algorithm with a signal to noise of 34. A second, more dedicated analysis shows this event is a glitch. We found a best-fitting solution with $\Delta\dot\nu<0$. This change of sign depends on factors such as the data span used for fitting the pre-glitch and post-glitch timing model, and the glitch epoch defined (not properly found by the algorithm). Therefore, it is essential to characterize glitches manually once they are detected by the automatic algorithm in order to derive more robust glitch parameters.

\begin{figure}[h!]
    \includegraphics[width=\linewidth]{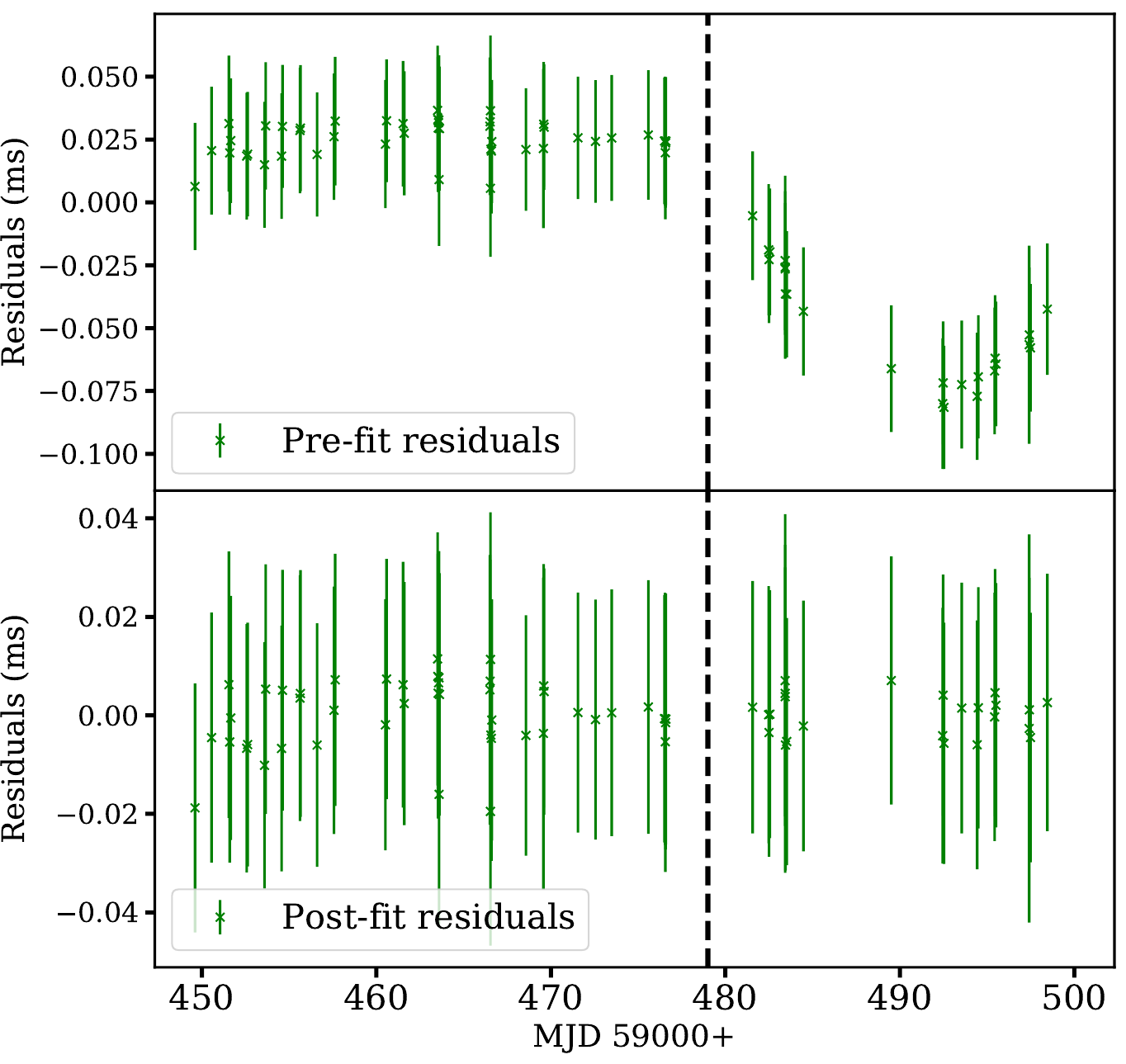}
    \caption{Detection of a small glitch 62 days after the 2021 giant glitch. \textit{Top:} Residuals without considering the small glitch. \textit{Bottom:} Residuals after fitting for the small glitch.}
    \label{fig: J0835-glitch2}
\end{figure}

We analysed the residuals restricted to a data span of $30~\mathrm{d}$ around the glitch in order to avoid effects of the short exponential recoveries of the giant glitch \citep{2023MNRAS.521.4504Z}. We show the associated residuals in Fig.~\ref{fig: J0835-glitch2}, and summarize these results in Table~\ref{tab:updated_parameters}.
We defined $t_\mathrm{g}$ halfway between the last pre-glitch observation and the first post-glitch observation.
Note that this glitch has a typical signature of $\Delta \nu >0$ and $\Delta \dot \nu <0$ but the magnitude of both of them is smaller than in giant glitches. This is the smallest glitch reported for the Vela pulsar so far. 

\subsection{PSR J1048--5832}
We detected a total of 14 events. It is seen from Fig.~\ref{fig: result} that the events are more concentrated in quadrants II and III  (nine events in total) than in I and IV (five events). Therefore, discrete events in this pulsar usually present $\Delta \nu<0$, while the sign of $\Delta \dot \nu$ is poorly constrained.
We also note that the events in this pulsar are the largest in our sample. They lie in the range $ 7\times10^{-9}<\lvert \Delta\nu (s^{-1}) \rvert < 3 \times 10^{-7}$ and $7\times10^{-16}<\lvert \Delta\dot\nu (s^{-2}) \rvert < 1.3 \times 10^{-13}$.

A more careful inspection revealed that the most significant events in quadrant II and III actually correspond to two small glitches occurring in 2022 and 2023.

\begin{figure}[h!]
    \includegraphics[width=\linewidth]{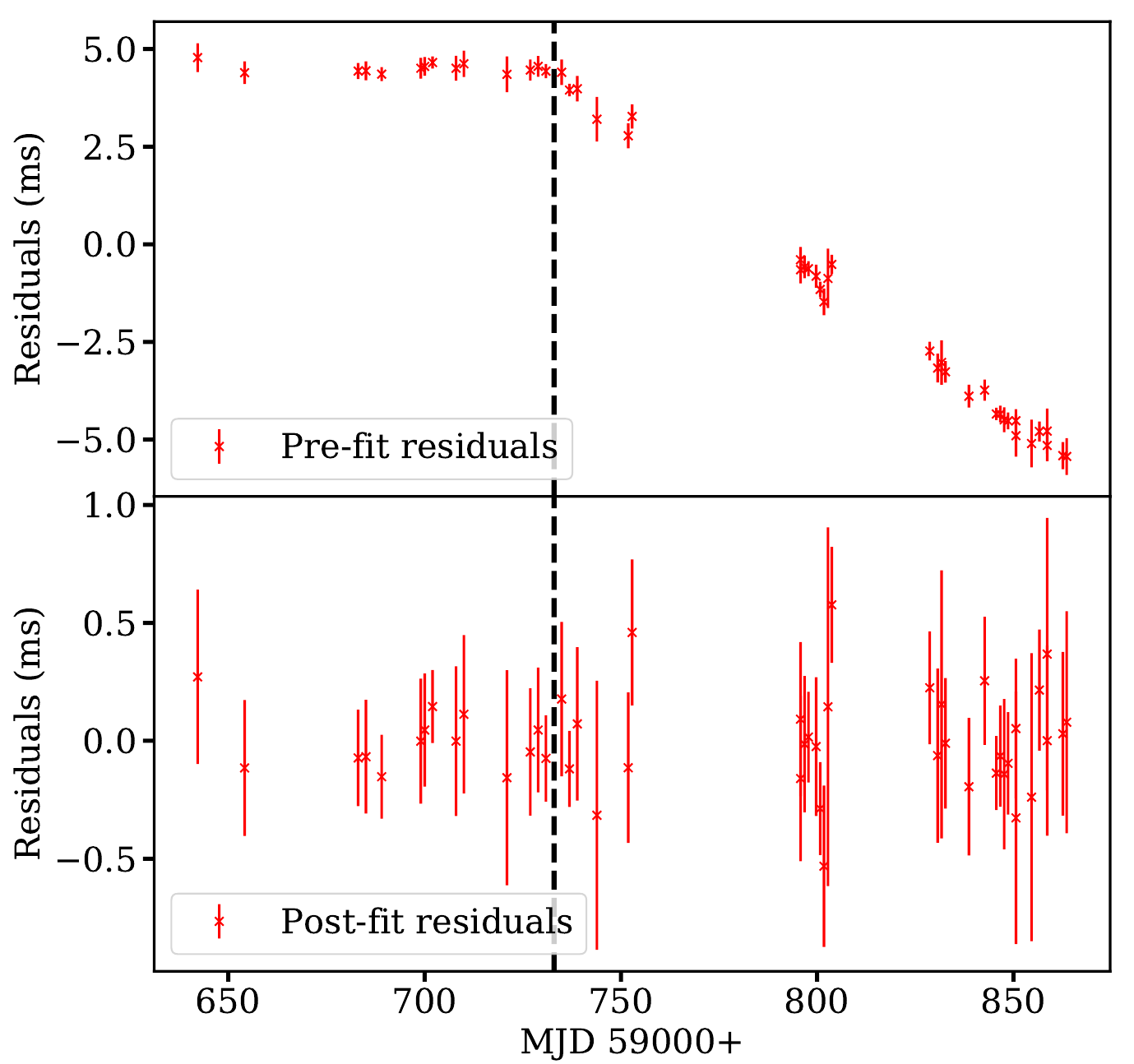}
    \caption{Glitch detected with the algorithm on MJD 59732.9(1) in PSR J1048$-$5832 (glitch 11). \textit{Top:} Residuals without considering the glitch. \textit{Bottom:} Residuals after fitting for the glitch.}
    \label{fig: J1048-glitch2}
\end{figure}

\begin{figure}[h!]
    \includegraphics[width=\linewidth]{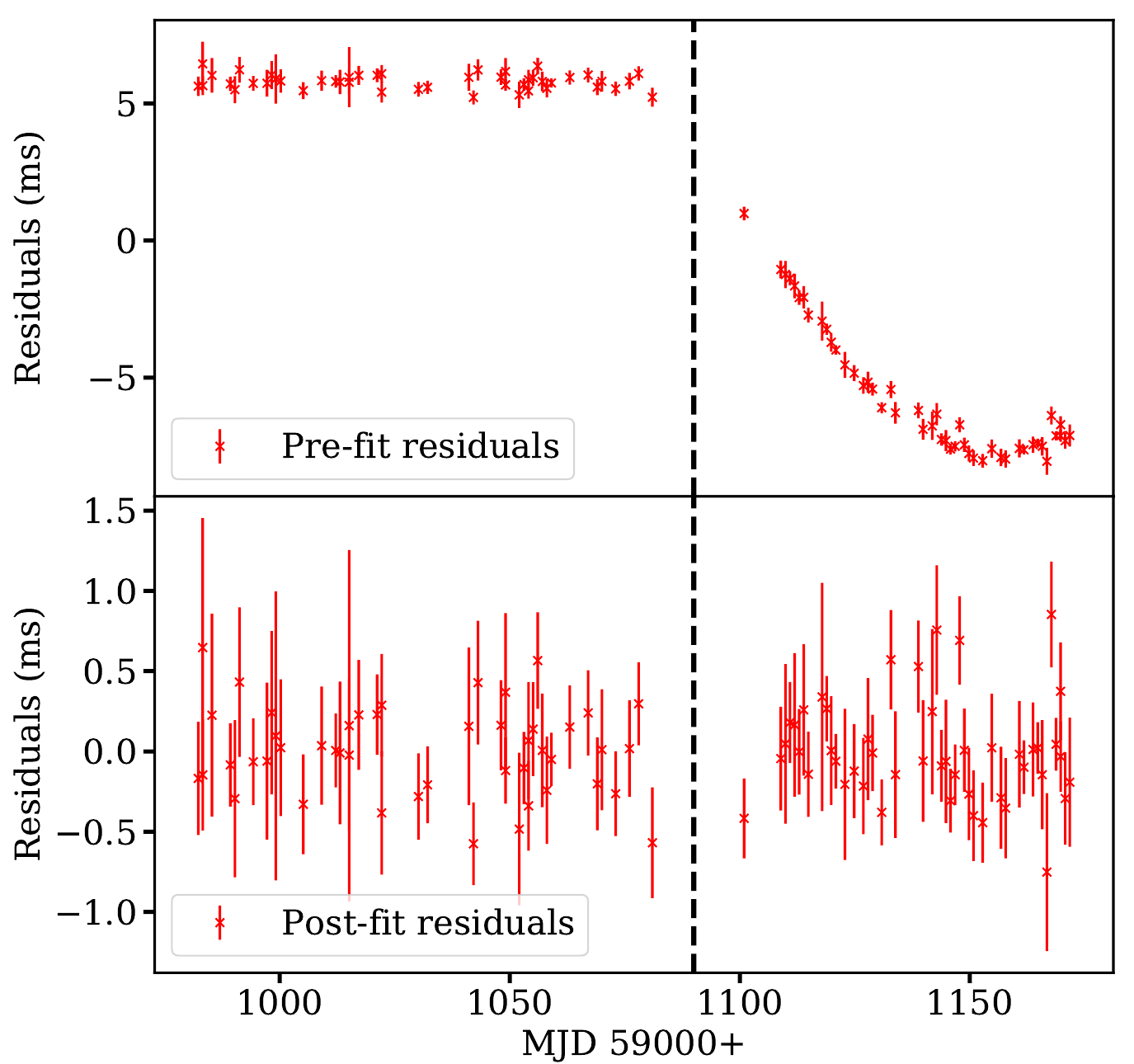}
    \caption{Glitch detected with the algorithm on MJD 60090(10) in PSR J1048$-$5832 (glitch 12). \textit{Top:} Residuals without considering the glitch. \textit{Bottom:} Residuals after fitting for the glitch.}
    \label{fig: J1048-glitch}
\end{figure}

In \cite{2023MNRAS.521.4504Z} we reported the detection of two small glitches for PSR~J1048--5832. Those two glitches are named as glitch 9 and glitch 10 in \cite{2023arXiv231204305L}. Our algorithm detected the first of them, being the event with the highest significance, but missed the second one, because our observation cadence around that epoch was too low compared to the minimum cadence required by the automatic process. In addition, the algorithm detected another 13 events, two of which were classified as glitches after a more thorough analysis. One of the new small glitches happened around MJD~59730 (glitch 11), and the other new one occurred around MJD~60090 (glitch 12). The last one can also be seen from the residuals in Fig.~\ref{fig: residuals1}. Although the algorithm detected these two glitches in quadrants II and III respectively, we found a best-fitting solution for both of them with $\Delta \nu > 0$ and $\Delta \dot \nu < 0$.
To characterize these new glitches (glitches 11 and 12 in this pulsar), we restricted the data span to $\sim$100 days around the glitch epoch (Figs. \ref{fig: J1048-glitch2}, \ref{fig: J1048-glitch}). The glitch parameters are given in Table \ref{tab:updated_parameters}.

We defined $t_\mathrm{g}$ as halfway between the last pre-glitch and the first post-glitch observation. Both glitches are smaller in amplitude than those reported in \cite{2023MNRAS.521.4504Z}. The two glitches reported in \cite{2023MNRAS.521.4504Z} correspond to the glitches 9 and 10 in \cite{2023arXiv231204305L}. Those glitches, together with glitch 8 reported in \cite{2023arXiv231204305L}, possess $\Delta\dot\nu_\mathrm{g}/\dot\nu < 0$. However, for these new two glitches (glitch 11 and glitch 12), we obtained $\Delta\dot\nu_\mathrm{g}/\dot\nu >0$, just like glitches 6 and 7 reported in \cite{2023arXiv231204305L}, which correspond to a more typical behaviour for glitches. 

In addition, while we were finishing this work, we detected \citep{2024ATel_J1048} a giant glitch in this pulsar (Table \ref{tab:updated_parameters}). A more thorough analysis of this glitch will be performed in a future work, once we obtain enough post-glitch TOAs from our ongoing monitoring program.


\subsection{PSR J1644--4559 and PSR J1731--4744}
We found 26 discrete events for PSR J1644$-$4559, 20 of which have $\Delta \nu > 0$, while for most of them the sign of $\Delta \dot \nu$ is unconstrained. The detections oscillate in a small range of $ 4\times10^{-10}<\lvert \Delta\nu (s^{-1}) \rvert < 1.7 \times 10^{-9}$ and $3\times10^{-17}<\lvert \Delta\dot\nu (s^{-2}) \rvert < 1.3 \times 10^{-15}$.

For PSR J1731$-$4744, the big uncertainties in its TOAs do not allow us to perform a thorough search for small glitches, and we can see that the five detections are close to or below the detection limits.

\subsection{PSR J1740--3015}

We detected 11 events: seven (63\%) correspond to $\Delta \nu < 0$. Three of them have $\Delta \dot\nu < 0$, while the sign of $\Delta \dot\nu$ is undefined in the rest of them. For the four events with $\Delta \nu > 0$, one of them possesses $\Delta \dot\nu < 0$, one $\Delta \dot\nu > 0$, while the sign of $\Delta \dot\nu$ is undefined in the other two. 
The events are in the range of $ 2.1\times10^{-9}<\lvert \Delta\nu (s^{-1}) \rvert < 2.1 \times 10^{-8}$ and $9\times10^{-17}<\lvert \Delta\dot\nu (s^{-2}) \rvert < 4.6 \times 10^{-15}$.

\begin{figure}[h!]
    \includegraphics[width=\linewidth]{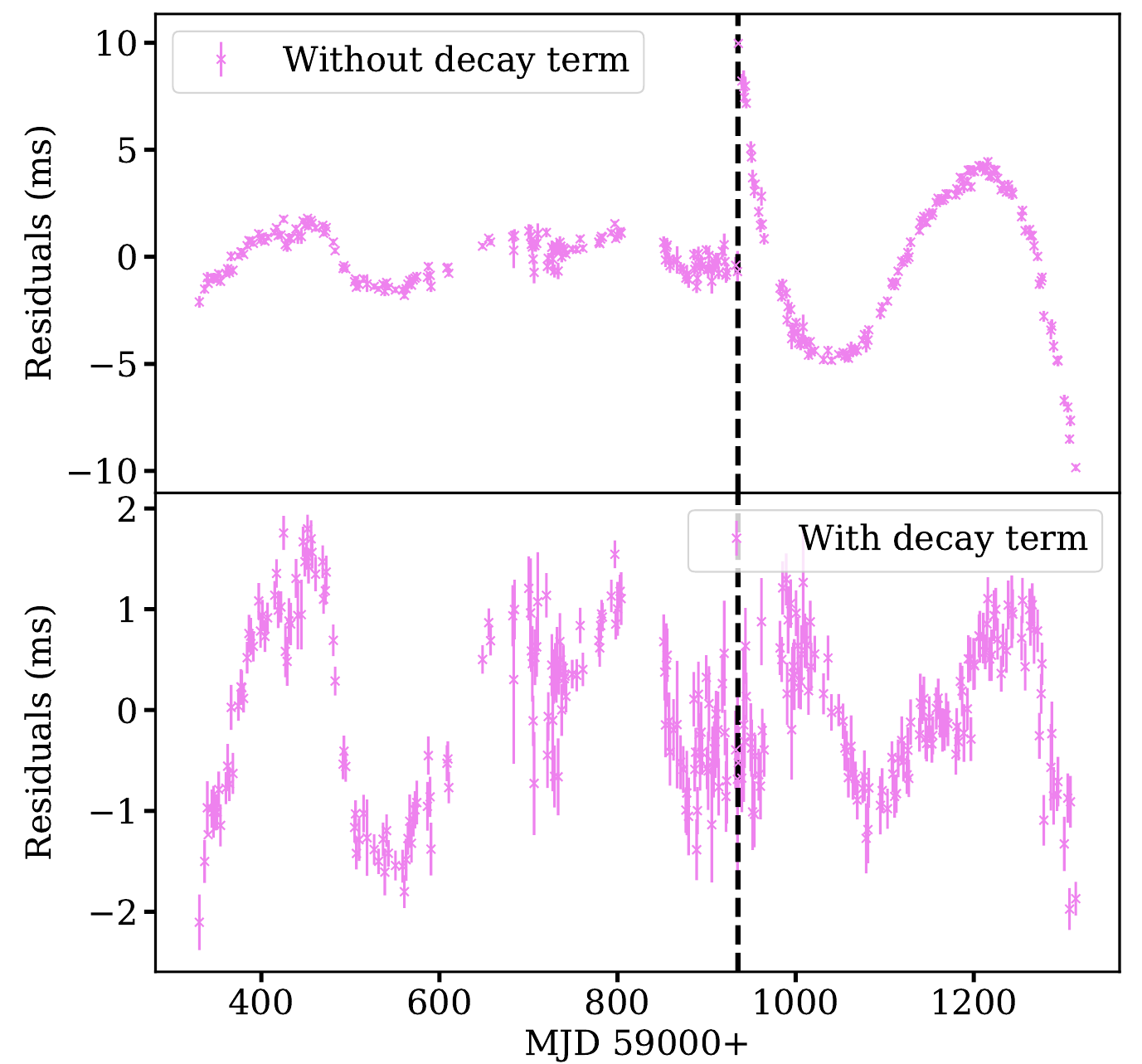}
    \caption{Detection of a new decay term in the PSR J1740$-$3015 glitch on MJD=59935.1(4). \textit{Top:} Residuals without considering any decay term. \textit{Bottom:} Residuals considering one decay term.}
    \label{fig: J1740-glitch}
\end{figure}

The algorithm also detected the 2022 glitch (MJD 59935.1(4)) reported in \cite{2023MNRAS.521.4504Z}, which was excluded from Fig. \ref{fig: result}. In this glitch, we detected a new long-scale decay term. We found
$\tau = 124(2)~\mathrm{d}$, with a degree of recovery of  $Q=4.43(0.01) \%$, which can be seen in Fig. \ref{fig: J1740-glitch}. We did not detect any change in $\Ddot{\nu}$. Updated parameters for the glitch are shown in Table \ref{tab:updated_parameters}.

\begin{table*}[htbp]
    \centering
    \caption{Updated parameters for all glitches detected by PuMA collaboration between MJD 58832 and MJD 60411. }
    \begin{tabular}{ccccccccc}
        \hline\hline
        PSR & Age  & Glitch epoch & $\Delta \nu_g / \nu$ & $\Delta \dot\nu_g / \dot\nu$ & $\Delta \Ddot\nu / \Ddot\nu$ & Q & $\tau_d$ & References \\
         & (kyr) & (MJD) & ($10^{-9}$) & ($10^{-3}$) & ($10^{-3}$) & & (d) & \\
        \hline
        J0742--2822 & $157$ & 59839.4(5) & 4295(1) & 72(1) & - & 0.446(1)\% & 33.4(5) & \cite{2023MNRAS.521.4504Z}\\ \hline
        J0835--4510 & $11$ & 59417.6194(2) & 1248(1) & 83(4) & -778.5(5)  & 0.7(1)\% & 0.994(8) & \cite{2023MNRAS.521.4504Z} \\
         &  &  & & &  & 0.2(1)\% & 6.400(2) & \\
         &  &  & & &  & 41(1)\% & 535(8) & \\
         &  & 59479.0(2.5) & 0.20(2) & 0.13(1) & - & & & This work \\ \hline
        J1048--5832 & $20$ & 59203.9(5) (\#9) & 8.90(9) & 0.62(1) & - & - &- & \cite{2023MNRAS.521.4504Z} \\
         &  & 59540(2) (\#10) & 9.9(3) & - & -  &- &-& \cite{2023MNRAS.521.4504Z}\\
         &  & 59732.9(1) (\#11) & 0.91(4) & 0.009(8) & - & -&-& This work\\
         &  & 60090(10) (\#12) & 4.5(1) & 1.01(4) & - & -&- & This work \\
         &  & 60406.6(5) (\#13)$^\dag$  & 4078(10) & 11.0(5) & - & -&- & \cite{2024ATel_J1048} \\ \hline
         J1740--3015 & $21$ & 59935.1(4) & 329.4(2) & 2.32(1) & - & 4.43(1)\% & 124(2) & \cite{2023MNRAS.521.4504Z} \\
        \hline
    \end{tabular}
    \label{tab:updated_parameters}
    \tablefoot{Numeration with (\#) in PSR J1048--5832 glitches follows \cite{2023arXiv231204305L}. $^\dag$ Very recent glitch.}
\end{table*}

%
\section{Discussion}\label{sec:discussion}

High-cadence pulsar observations are essential to characterize small glitches and short decay components in giant glitches \citep{2014MNRAS.440.2755E, 2020MNRAS.491.3182B, 2021MNRAS.505.5488S, 2023MNRAS.521.4504Z}, and to better understand the distinction between red noise and glitches.
Considering that we detected one small glitch in PSR~J1048$-$5832 and one in the Vela pulsar that were not detected by visual inspection of the residuals, this study highlights the importance of developing automated detection codes for systematically identifying glitches \citep{2020ApJ...896...78M, 2021MNRAS.505.5488S}. 
In addition, it is important to follow up each event with a more thorough analysis in order to classify them as glitches or timing irregularities, and in the former case also characterise the glitch parameters. It would be interesting to develop a more robust algorithm also capable of disentangling between small glitches and timing irregularities.

Such an approach is essential for accurately discerning the contribution of glitches to red noise and distinguishing it from the inherent smooth wandering around a stable rotational model, enhancing our understanding of underlying phenomena for red noise and glitch mechanisms.

We have identified irregularities in all the six glitching pulsars that we studied, with different minimum and maximum sizes in $\lvert \Delta \nu \rvert$ and $\lvert \Delta \dot \nu \rvert$. In the particular case of the Vela pulsar, most of the events have a positive change in $\dot\nu$, i.e. are concentrated in the I and II quadrants. Using a similar systematic search over more than 20 yr of data, \cite{2021A&A...647A..25E} found that the numbers of events in all quadrants are comparable, which is different from what we obtain. The only asymmetry they found is that the size distributions of the events in the second and fourth quadrants do not appear to come from the same parent distribution. 

For the rest of the pulsars, events appear to be randomly distributed in the four quadrants. This resembles the result of \citet{2021A&A...647A..25E}, demonstrating that rotation of pulsars are affected by small variations in $\nu$ and $\dot\nu$. Considering that only three of these events proved to be glitches, we conclude that only a small portion of the timing noise seen in pulsars can be accounted for by small glitches. Therefore, our result indicates the presence of timing irregularities distinct from glitches, implying the existence of additional processes or phenomena that are not fully understood yet. A similar conclusion was reached by \cite{1988ApJ...330..847C} and \cite{2021A&A...647A..25E} for the Vela pulsar. Besides these small irregularities were sometimes presented as glitches or anti-glitches \citep[see, i.e,][and references therein]{2024ApJ...967L..13T}, we have shown that those irregularities are instead quite frequent, showing also a rather slower evolution compared to glitches, which are characterized by sudden significant changes in $\nu$ and $\dot{\nu}$, that only take place sporadically.

Regarding the exploration of red noise phenomena, we have discovered that, for PSR~J0742$-$2822, the red noise diminishes significantly after the glitch event on MJD 59839.4(5) firstly detected by \cite{2022ATel15629....1G}.
This behaviour resembles what was reported by \cite{2013MNRAS.432.3080K} for the 2009 glitch in the same pulsar. \cite{1990MNRAS.246..364J} proposed that the red noise is influenced by the interaction between superfluid neutron vortices and the Coulomb lattice at the solid crust. In this scenario, there is a possible correlation between red noise and the presence of vortices that are pinned to the pulsar crust, with the red noise diminishing after vortices are dragged away during the glitch. Our results for J0742$-$2822 are also consistent with this interpretation. 

In contrast, in the cases of PSR J1740$-$3015 and J0835$-$4510, we did not observe a significant decrease in red noise following the giant glitches on MJD 59935.1 and MJD 59417.6193, respectively. Considering that the properties of timing noise vary across the pulsar population \citep{2010ApJ...725.1607S}, and that the characteristic age of PSR J0742$-$2822 is $\sim 10^5~\mathrm{yr}$, whereas for PSR J0835$-$4510 and PSR J1740$-$3015 the characteristic age is a factor $\sim10$ smaller ($\sim 10^4~\mathrm{yr}$) \citep{2005yCat.7245....0M}, our result indicates that not only the timing noise varies across the pulsar population but also the entanglement between red noise and glitches changes across the pulsar population. In addition, our result supports the theory that the red noise may come from both internal and external torque \citep{2010Sci...329..408L,2014MNRAS.437...21M}, and that the change in red noise behaviour during a glitch depends on the particular origin of red noise in each case. Another possibility is that this difference in red noise behaviour during a glitch could indicate the existence of different forms of turbulence in the superfluid interior of neutron stars \citep{2014MNRAS.437...21M}.

We also found a small glitch in the Vela pulsar reported in Table \ref{tab:updated_parameters}. This is 62 days after the 2021 giant glitch.
 \cite{2021A&A...647A..25E} and \cite{2023MNRAS.522.5469D} reported one small glitch each in the Vela pulsar, 93 and 179 days after its giant glitch in 1991. \cite{2021A&A...647A..25E} reported another small glitch 134 days before the 2000 giant glitch. 
\cite{2020MNRAS.494..228L} reported one small glitch 379 days after the 2013 giant glitch.
In the case of the Vela pulsar, large glitches occur approximately every $\sim$3 years. However, small glitches mentioned here do not seem to follow any pattern in their inter-glitch times. It is probable that many small glitches in the Vela pulsar have gone unnoticed due to the lack of permanent high-cadence monitoring. More detections of small glitches in frequently glitching pulsars are required to further investigate the underlying mechanisms and relationships between these events.

%

%
\section{Conclusions} \label{sec:conclusions}
%

We presented a detailed analysis of the IAR pulsar monitoring campaign between 2019 and 2023 and developed an algorithm to look for small glitches in timing data. We found in our data span (of almost four years) three new small glitches, two in PSR J1048$-$5832 and one in Vela pulsar, that were not detected by visual inspection. This demonstrates the necessity of developing algorithms to look systematically for small glitches and not only rely on visual inspection of residuals. Also, high-cadence monitoring is required to further study population of small glitches, in order to gain knowledge on glitches mechanisms.

In addition to the small glitches detected, we found many irregularities that could not be associated with glitches, which suggest the presence of a mechanism for red noise inherently different from the mechanisms that produce glitches. However, we showed that for the 2022 glitch in PSR J0742$-$2822 red noise diminished significantly after the glitch. Thus it can not be assumed that red noise is a completely independent phenomenon from all glitches.


\begin{acknowledgements}
      FG and JAC are CONICET researchers and acknowledge support by PIP 0113 (CONICET). FG acknowledges support by PIBAA 1275 (CONICET). COL gratefully acknowledges the National Science Foundation (NSF) for financial support from Grant No. PHY-2207920. CME acknowledges support from the grant ANID FONDECYT 1211964. We also extend our gratitude to the technical staff at the IAR for their continuous efforts that enable our high-cadence observational campaign. 
      JAC was also supported by grant PID2022-136828NB-C42 funded by the Spanish MCIN/AEI/ 10.13039/501100011033 and “ERDF A way of making Europe” and by Consejería de Economía, Innovación, Ciencia y Empleo of Junta de Andalucía as research group FQM-322.

\end{acknowledgements}

%
\bibliographystyle{aa} 
\bibliography{biblio} 

\begin{thebibliography}{62}
\expandafter\ifx\csname natexlab\endcsname\relax\def\natexlab#1{#1}\fi

\bibitem[{{Antonelli} {et~al.}(2023){Antonelli}, {Basu}, \& {Haskell}}]{2023MNRAS.520.2813A}
{Antonelli}, M., {Basu}, A., \& {Haskell}, B. 2023, \mnras, 520, 2813

\bibitem[{{Antonopoulou} {et~al.}(2022){Antonopoulou}, {Haskell}, \& {Espinoza}}]{2022RPPh...85l6901A}
{Antonopoulou}, D., {Haskell}, B., \& {Espinoza}, C.~M. 2022, Reports on Progress in Physics, 85, 126901

\bibitem[{{Araujo Furlan} {et~al.}(2024){Araujo Furlan}, {Zubieta}, {Gancio}, {Romero}, {del Palacio}, {Garc{\'\i}a}, {Lousto}, \& {Combi}}]{2024RMxAC..56...85A}
{Araujo Furlan}, S.~B., {Zubieta}, E., {Gancio}, G., {et~al.} 2024, in Revista Mexicana de Astronomia y Astrofisica Conference Series, Vol.~56, Revista Mexicana de Astronomia y Astrofisica Conference Series, 85--89

\bibitem[{{Arzoumanian} {et~al.}(1994){Arzoumanian}, {Nice}, {Taylor}, \& {Thorsett}}]{1994ApJ...422..671A}
{Arzoumanian}, Z., {Nice}, D.~J., {Taylor}, J.~H., \& {Thorsett}, S.~E. 1994, \apj, 422, 671

\bibitem[{{Basu} {et~al.}(2020){Basu}, {Joshi}, {Krishnakumar}, {Bhattacharya}, {Nandi}, {Bandhopadhay}, {Char}, \& {Manoharan}}]{2020MNRAS.491.3182B}
{Basu}, A., {Joshi}, B.~C., {Krishnakumar}, M.~A., {et~al.} 2020, \mnras, 491, 3182

\bibitem[{{Basu} {et~al.}(2022){Basu}, {Shaw}, {Antonopoulou}, {Keith}, {Lyne}, {Mickaliger}, {Stappers}, {Weltevrede}, \& {Jordan}}]{2022MNRAS.510.4049B}
{Basu}, A., {Shaw}, B., {Antonopoulou}, D., {et~al.} 2022, \mnras, 510, 4049

\bibitem[{{Baym} {et~al.}(1969){Baym}, {Pethick}, \& {Pines}}]{1969Natur.224..673B}
{Baym}, G., {Pethick}, C., \& {Pines}, D. 1969, \nat, 224, 673

\bibitem[{{Cordes} {et~al.}(1988){Cordes}, {Downs}, \& {Krause-Polstorff}}]{1988ApJ...330..847C}
{Cordes}, J.~M., {Downs}, G.~S., \& {Krause-Polstorff}, J. 1988, \apj, 330, 847

\bibitem[{{Dunn} {et~al.}(2023){Dunn}, {Melatos}, {Espinoza}, {Antonopoulou}, \& {Dodson}}]{2023MNRAS.522.5469D}
{Dunn}, L., {Melatos}, A., {Espinoza}, C.~M., {Antonopoulou}, D., \& {Dodson}, R. 2023, \mnras, 522, 5469

\bibitem[{{Espinoza} {et~al.}(2021){Espinoza}, {Antonopoulou}, {Dodson}, {Stepanova}, \& {Scherer}}]{2021A&A...647A..25E}
{Espinoza}, C.~M., {Antonopoulou}, D., {Dodson}, R., {Stepanova}, M., \& {Scherer}, A. 2021, \aap, 647, A25

\bibitem[{{Espinoza} {et~al.}(2014){Espinoza}, {Antonopoulou}, {Stappers}, {Watts}, \& {Lyne}}]{2014MNRAS.440.2755E}
{Espinoza}, C.~M., {Antonopoulou}, D., {Stappers}, B.~W., {Watts}, A., \& {Lyne}, A.~G. 2014, \mnras, 440, 2755

\bibitem[{{Espinoza} {et~al.}(2011){Espinoza}, {Lyne}, {Stappers}, \& {Kramer}}]{2011MNRAS.414.1679E}
{Espinoza}, C.~M., {Lyne}, A.~G., {Stappers}, B.~W., \& {Kramer}, M. 2011, \mnras, 414, 1679

\bibitem[{{Fuentes} {et~al.}(2017){Fuentes}, {Espinoza}, {Reisenegger}, {Shaw}, {Stappers}, \& {Lyne}}]{2017A&A...608A.131F}
{Fuentes}, J.~R., {Espinoza}, C.~M., {Reisenegger}, A., {et~al.} 2017, \aap, 608, A131

\bibitem[{{Gancio} {et~al.}(2020){Gancio}, {Lousto}, {Combi}, {del Palacio}, {L{\'o}pez Armengol}, {Combi}, {Garc{\'\i}a}, {Kornecki}, {M{\"u}ller}, {Guti{\'e}rrez}, {Hauscarriaga}, \& {Mancuso}}]{2020A&A...633A..84G}
{Gancio}, G., {Lousto}, C.~O., {Combi}, L., {et~al.} 2020, \aap, 633, A84

\bibitem[{{Grover} {et~al.}(2024){Grover}, {Joshi}, {Singha}, {G{\"u}gercino{\u{g}}lu}, {Arumugam}, {Bandyopadhyay}, {Chibueze}, {Desai}, {Eya}, {Kundu}, \& {Urama}}]{2024arXiv240514351G}
{Grover}, H., {Joshi}, B.~C., {Singha}, J., {et~al.} 2024, arXiv e-prints, arXiv:2405.14351

\bibitem[{{Grover} {et~al.}(2022){Grover}, {Singha}, {Joshi}, \& {Arumugam}}]{2022ATel15629....1G}
{Grover}, H., {Singha}, J., {Joshi}, B.~C., \& {Arumugam}, P. 2022, The Astronomer's Telegram, 15629, 1

\bibitem[{{G{\"u}gercino{\u{g}}lu} {et~al.}(2022){G{\"u}gercino{\u{g}}lu}, {Ge}, {Yuan}, \& {Zhou}}]{2022MNRAS.511..425G}
{G{\"u}gercino{\u{g}}lu}, E., {Ge}, M.~Y., {Yuan}, J.~P., \& {Zhou}, S.~Q. 2022, \mnras, 511, 425

\bibitem[{{Haskell} \& {Melatos}(2015)}]{2015IJMPD..2430008H}
{Haskell}, B. \& {Melatos}, A. 2015, International Journal of Modern Physics D, 24, 1530008

\bibitem[{{Hobbs} {et~al.}(2012){Hobbs}, {Coles}, {Manchester}, {Keith}, {Shannon}, {Chen}, {Bailes}, {Bhat}, {Burke-Spolaor}, {Champion}, {Chaudhary}, {Hotan}, {Khoo}, {Kocz}, {Levin}, {Oslowski}, {Preisig}, {Ravi}, {Reynolds}, {Sarkissian}, {van Straten}, {Verbiest}, {Yardley}, \& {You}}]{2012MNRAS.427.2780H}
{Hobbs}, G., {Coles}, W., {Manchester}, R.~N., {et~al.} 2012, \mnras, 427, 2780

\bibitem[{{Hobbs} {et~al.}(2010){Hobbs}, {Lyne}, \& {Kramer}}]{2010MNRAS.402.1027H}
{Hobbs}, G., {Lyne}, A.~G., \& {Kramer}, M. 2010, \mnras, 402, 1027

\bibitem[{{Hobbs} {et~al.}(2006){Hobbs}, {Edwards}, \& {Manchester}}]{2006MNRAS.369..655H}
{Hobbs}, G.~B., {Edwards}, R.~T., \& {Manchester}, R.~N. 2006, \mnras, 369, 655

\bibitem[{{Hotan} {et~al.}(2004){Hotan}, {van Straten}, \& {Manchester}}]{2004PASA...21..302H}
{Hotan}, A.~W., {van Straten}, W., \& {Manchester}, R.~N. 2004, \pasa, 21, 302

\bibitem[{{Janssen} \& {Stappers}(2006)}]{2006A&A...457..611J}
{Janssen}, G.~H. \& {Stappers}, B.~W. 2006, \aap, 457, 611

\bibitem[{{Jones}(1990)}]{1990MNRAS.246..364J}
{Jones}, P.~B. 1990, \mnras, 246, 364

\bibitem[{{Keith} {et~al.}(2013){Keith}, {Shannon}, \& {Johnston}}]{2013MNRAS.432.3080K}
{Keith}, M.~J., {Shannon}, R.~M., \& {Johnston}, S. 2013, \mnras, 432, 3080

\bibitem[{{Lentati} {et~al.}(2014){Lentati}, {Alexander}, {Hobson}, {Feroz}, {van Haasteren}, {Lee}, \& {Shannon}}]{2014MNRAS.437.3004L}
{Lentati}, L., {Alexander}, P., {Hobson}, M.~P., {et~al.} 2014, \mnras, 437, 3004

\bibitem[{{Liu} {et~al.}(2023){Liu}, {Yuan}, {Ge}, {Ye}, {Zhou}, {Dang}, {Zhou}, {G{\"u}gercino{\u{g}}lu}, {Wang}, {Wang}, {Li}, {Li}, \& {Wang}}]{2023arXiv231204305L}
{Liu}, P., {Yuan}, J.~P., {Ge}, M.~Y., {et~al.} 2023, arXiv e-prints, arXiv:2312.04305

\bibitem[{{Lopez Armengol} {et~al.}(2019){Lopez Armengol}, {Lousto}, {del Palacio}, {Garcia}, {Combi}, {Combi}, {Gancio}, {Mueller}, \& {Kornecki}}]{2019ATel12482....1L}
{Lopez Armengol}, F.~G., {Lousto}, C.~O., {del Palacio}, S., {et~al.} 2019, The Astronomer's Telegram, 12482, 1

\bibitem[{{Lorimer} \& {Kramer}(2004)}]{2004hpa..book.....L}
{Lorimer}, D.~R. \& {Kramer}, M. 2004, {Handbook of Pulsar Astronomy}, Vol.~4

\bibitem[{{Lousto} {et~al.}(2022){Lousto}, {Missel}, {Prajapati}, {Sosa Fiscella}, {Armengol}, {Gyawali}, {Wang}, {Cahill}, {Combi}, {Palacio}, {Combi}, {Gancio}, {Garc{\'\i}a}, {Guti{\'e}rrez}, \& {Hauscarriaga}}]{2022MNRAS.509.5790L}
{Lousto}, C.~O., {Missel}, R., {Prajapati}, H., {et~al.} 2022, \mnras, 509, 5790

\bibitem[{{Lousto} {et~al.}(2024){Lousto}, {Missel}, {Zubieta}, {del Palacio}, {Garc{\'\i}a}, {Gancio}, {Wang}, {Araujo Furlan}, \& {Combi}}]{2024RMxAC..56..134L}
{Lousto}, C.~O., {Missel}, R., {Zubieta}, E., {et~al.} 2024, in Revista Mexicana de Astronomia y Astrofisica Conference Series, Vol.~56, Revista Mexicana de Astronomia y Astrofisica Conference Series, 134--144

\bibitem[{{Lower} {et~al.}(2020){Lower}, {Bailes}, {Shannon}, {Johnston}, {Flynn}, {Os{\l}owski}, {Gupta}, {Farah}, {Bateman}, {Green}, {Hunstead}, {Jameson}, {Jankowski}, {Parthasarathy}, {Price}, {Sutherland}, {Temby}, \& {Venkatraman Krishnan}}]{2020MNRAS.494..228L}
{Lower}, M.~E., {Bailes}, M., {Shannon}, R.~M., {et~al.} 2020, \mnras, 494, 228

\bibitem[{{Luo} {et~al.}(2021){Luo}, {Ransom}, {Demorest}, {Ray}, {Archibald}, {Kerr}, {Jennings}, {Bachetti}, {van Haasteren}, {Champagne}, {Colen}, {Phillips}, {Zimmerman}, {Stovall}, {Lam}, \& {Jenet}}]{2021ApJ...911...45L}
{Luo}, J., {Ransom}, S., {Demorest}, P., {et~al.} 2021, \apj, 911, 45

\bibitem[{{Luo} {et~al.}(2019){Luo}, {Ransom}, {Demorest}, {van Haasteren}, {Ray}, {Stovall}, {Bachetti}, {Archibald}, {Kerr}, {Colen}, \& {Jenet}}]{2019ascl.soft02007L}
{Luo}, J., {Ransom}, S., {Demorest}, P., {et~al.} 2019, {PINT: High-precision pulsar timing analysis package}, Astrophysics Source Code Library, record ascl:1902.007

\bibitem[{{Lyne} {et~al.}(2010){Lyne}, {Hobbs}, {Kramer}, {Stairs}, \& {Stappers}}]{2010Sci...329..408L}
{Lyne}, A., {Hobbs}, G., {Kramer}, M., {Stairs}, I., \& {Stappers}, B. 2010, Science, 329, 408

\bibitem[{{Manchester}(2018)}]{2018IAUS..337..197M}
{Manchester}, R.~N. 2018, in Pulsar Astrophysics the Next Fifty Years, ed. P.~{Weltevrede}, B.~B.~P. {Perera}, L.~L. {Preston}, \& S.~{Sanidas}, Vol. 337, 197--202

\bibitem[{{Manchester} {et~al.}(2005{\natexlab{a}}){Manchester}, {Hobbs}, {Teoh}, \& {Hobbs}}]{2005AJ....129.1993M}
{Manchester}, R.~N., {Hobbs}, G.~B., {Teoh}, A., \& {Hobbs}, M. 2005{\natexlab{a}}, \aj, 129, 1993

\bibitem[{{Manchester} {et~al.}(2005{\natexlab{b}}){Manchester}, {Hobbs}, {Teoh}, \& {Hobbs}}]{2005yCat.7245....0M}
{Manchester}, R.~N., {Hobbs}, G.~B., {Teoh}, A., \& {Hobbs}, M. 2005{\natexlab{b}}, VizieR Online Data Catalog, VII/245

\bibitem[{{Melatos} {et~al.}(2020){Melatos}, {Dunn}, {Suvorova}, {Moran}, \& {Evans}}]{2020ApJ...896...78M}
{Melatos}, A., {Dunn}, L.~M., {Suvorova}, S., {Moran}, W., \& {Evans}, R.~J. 2020, \apj, 896, 78

\bibitem[{{Melatos} \& {Link}(2014)}]{2014MNRAS.437...21M}
{Melatos}, A. \& {Link}, B. 2014, \mnras, 437, 21

\bibitem[{{Melatos} \& {Warszawski}(2009)}]{2009ApJ...700.1524M}
{Melatos}, A. \& {Warszawski}, L. 2009, \apj, 700, 1524

\bibitem[{{Parthasarathy} {et~al.}(2019){Parthasarathy}, {Shannon}, {Johnston}, {Lentati}, {Bailes}, {Dai}, {Kerr}, {Manchester}, {Os{\l}owski}, {Sobey}, {van Straten}, \& {Weltevrede}}]{2019MNRAS.489.3810P}
{Parthasarathy}, A., {Shannon}, R.~M., {Johnston}, S., {et~al.} 2019, \mnras, 489, 3810

\bibitem[{{Ransom}(2011)}]{2011ascl.soft07017R}
{Ransom}, S. 2011, {PRESTO: PulsaR Exploration and Search TOolkit}, Astrophysics Source Code Library, record ascl:1107.017

\bibitem[{{Ransom} {et~al.}(2003){Ransom}, {Cordes}, \& {Eikenberry}}]{2003ApJ...589..911R}
{Ransom}, S.~M., {Cordes}, J.~M., \& {Eikenberry}, S.~S. 2003, \apj, 589, 911

\bibitem[{{Shannon} \& {Cordes}(2010)}]{2010ApJ...725.1607S}
{Shannon}, R.~M. \& {Cordes}, J.~M. 2010, \apj, 725, 1607

\bibitem[{{Shannon} {et~al.}(2016){Shannon}, {Lentati}, {Kerr}, {Johnston}, {Hobbs}, \& {Manchester}}]{2016MNRAS.459.3104S}
{Shannon}, R.~M., {Lentati}, L.~T., {Kerr}, M., {et~al.} 2016, \mnras, 459, 3104

\bibitem[{{Shaw} {et~al.}(2022){Shaw}, {Mickaliger}, {Stappers}, {Lyne}, {Keith}, {Weltevrede}, \& {Basu}}]{2022ATel15622....1S}
{Shaw}, B., {Mickaliger}, M.~B., {Stappers}, B.~W., {et~al.} 2022, The Astronomer's Telegram, 15622, 1

\bibitem[{{Singha} {et~al.}(2021){Singha}, {Basu}, {Krishnakumar}, {Joshi}, \& {Arumugam}}]{2021MNRAS.505.5488S}
{Singha}, J., {Basu}, A., {Krishnakumar}, M.~A., {Joshi}, B.~C., \& {Arumugam}, P. 2021, \mnras, 505, 5488

\bibitem[{{Sosa Fiscella} {et~al.}(2021){Sosa Fiscella}, {del Palacio}, {Combi}, {Lousto}, {Combi}, {Gancio}, {Garc{\'\i}a}, {Guti{\'e}rrez}, {Hauscarriaga}, {Kornecki}, {L{\'o}pez Armengol}, {Mancuso}, {M{\"u}ller}, \& {Simaz Bunzel}}]{2021ApJ...908..158S}
{Sosa Fiscella}, V., {del Palacio}, S., {Combi}, L., {et~al.} 2021, \apj, 908, 158

\bibitem[{{Sosa-Fiscella} {et~al.}(2021){Sosa-Fiscella}, {Zubieta}, {del Palacio}, {Garcia}, {Lopez-Armengol}, {Combi}, {Lousto}, {Gancio}, {Combi}, {Gutierrez}, {Bunzel}, {Hauscarriaga}, \& {PuMA Collaboration}}]{2021ATel14806....1S}
{Sosa-Fiscella}, V., {Zubieta}, E., {del Palacio}, S., {et~al.} 2021, The Astronomer's Telegram, 14806, 1

\bibitem[{{Taylor}(1992)}]{1992RSPTA.341..117T}
{Taylor}, J.~H. 1992, Philosophical Transactions of the Royal Society of London Series A, 341, 117

\bibitem[{{Tuo} {et~al.}(2024){Tuo}, {Serim}, {Antonelli}, {Ducci}, {Vahdat}, {Ge}, {Santangelo}, \& {Xie}}]{2024ApJ...967L..13T}
{Tuo}, Y., {Serim}, M.~M., {Antonelli}, M., {et~al.} 2024, \apjl, 967, L13

\bibitem[{{Yu} {et~al.}(2013){Yu}, {Manchester}, {Hobbs}, {Johnston}, {Kaspi}, {Keith}, {Lyne}, {Qiao}, {Ravi}, {Sarkissian}, {Shannon}, \& {Xu}}]{2013MNRAS.429..688Y}
{Yu}, M., {Manchester}, R.~N., {Hobbs}, G., {et~al.} 2013, \mnras, 429, 688

\bibitem[{{Zhou} {et~al.}(2022{\natexlab{a}}){Zhou}, {G{\"u}gercino{\u{g}}lu}, {Yuan}, {Ge}, \& {Yu}}]{2022Univ....8..641Z}
{Zhou}, S., {G{\"u}gercino{\u{g}}lu}, E., {Yuan}, J., {Ge}, M., \& {Yu}, C. 2022{\natexlab{a}}, Universe, 8, 641

\bibitem[{{Zhou} {et~al.}(2022{\natexlab{b}}){Zhou}, {Wang}, {Wang}, {Yuan}, {Kou}, \& {Dang}}]{Zhou2022}
{Zhou}, Z.-R., {Wang}, J.-B., {Wang}, N., {et~al.} 2022{\natexlab{b}}, Research in Astronomy and Astrophysics, 22, 095008

\bibitem[{{Zubieta} {et~al.}(2024{\natexlab{a}}){Zubieta}, {Araujo Furlan}, {del Palacio}, {Garcia}, {Gancio}, {Lousto}, \& {Combi}}]{2024ATel16580....1Z}
{Zubieta}, E., {Araujo Furlan}, S.~B., {del Palacio}, S., {et~al.} 2024{\natexlab{a}}, The Astronomer's Telegram, 16580, 1

\bibitem[{{Zubieta} {et~al.}(2023{\natexlab{a}}){Zubieta}, {del Palacio}, {Garc{\'\i}a}, {Araujo Furlan}, {Gancio}, {Lousto}, \& {Combi}}]{2023arXiv231208188Z}
{Zubieta}, E., {del Palacio}, S., {Garc{\'\i}a}, F., {et~al.} 2023{\natexlab{a}}, arXiv e-prints, arXiv:2312.08188

\bibitem[{{Zubieta} {et~al.}(2024{\natexlab{b}}){Zubieta}, {del Palacio}, {Garc{\'\i}a}, {Araujo Furlan}, {Gancio}, {Lousto}, \& {Combi}}]{2024RMxAC..56..161Z}
{Zubieta}, E., {del Palacio}, S., {Garc{\'\i}a}, F., {et~al.} 2024{\natexlab{b}}, in Revista Mexicana de Astronomia y Astrofisica Conference Series, Vol.~56, Revista Mexicana de Astronomia y Astrofisica Conference Series, 161--165

\bibitem[{{Zubieta} {et~al.}(2022{\natexlab{a}}){Zubieta}, {Del Palacio}, {Garcia}, {Gancio}, {Lousto}, {Combi}, {Combi}, {Gutierrez}, {Lopez-Armengol}, {Simaz Bunzel}, \& {Sosa-Fiscella}}]{2022ATel15638....1Z}
{Zubieta}, E., {Del Palacio}, S., {Garcia}, F., {et~al.} 2022{\natexlab{a}}, The Astronomer's Telegram, 15638, 1

\bibitem[{{Zubieta} {et~al.}(2024{\natexlab{c}}){Zubieta}, {Furlan}, {del Palacio}, {Garcia}, {Gancio}, {Lousto}, {Combi}, \& {Combi}}]{2024ATel_J1048}
{Zubieta}, E., {Furlan}, S.~B.~A., {del Palacio}, S., {et~al.} 2024{\natexlab{c}}, The Astronomer's Telegram, 16580, 1

\bibitem[{{Zubieta} {et~al.}(2022{\natexlab{b}}){Zubieta}, {Furlan}, {Palacio}, {Garcia}, {Gancio}, {Lousto}, {Combi}, \& {Combi}}]{2022ATel15838....1Z}
{Zubieta}, E., {Furlan}, S.~B.~A., {Palacio}, S.~d., {et~al.} 2022{\natexlab{b}}, The Astronomer's Telegram, 15838, 1

\bibitem[{{Zubieta} {et~al.}(2023{\natexlab{b}}){Zubieta}, {Missel}, {Sosa Fiscella}, {Lousto}, {del Palacio}, {L{\'o}pez Armengol}, {Garc{\'\i}a}, {Combi}, {Wang}, {Combi}, {Gancio}, {Negrelli}, \& {Guti{\'e}rrez}}]{2023MNRAS.521.4504Z}
{Zubieta}, E., {Missel}, R., {Sosa Fiscella}, V., {et~al.} 2023{\natexlab{b}}, \mnras, 521, 4504

\end{thebibliography}
%


\end{document}